\documentclass[%
reprint,
superscriptaddress,
showpacs,preprintnumbers,
verbatim,
amsmath,amssymb,
aps,
pre,
]{revtex4-1}

\usepackage{color}
\usepackage{graphicx}
\usepackage{dcolumn}
\usepackage{bm}
\usepackage{textcomp} 

\begin {document}

\title{%
  Brownian motion with alternately fluctuating diffusivity:\\
  Stretched-exponential and power-law relaxation
}


\author{Tomoshige Miyaguchi}
\email{tmiyaguchi@naruto-u.ac.jp}
\affiliation{%
  Department of Mathematics, Naruto University of Education,
  Naruto, Tokushima 772-8502, Japan}

\author{Takashi Uneyama}
\affiliation{%
  Center for Computational Science, Graduate School of Engineering,
  Nagoya University, Furo-cho, Chikusa, Nagoya 464-8603,
  Japan
}%

\author{Takuma Akimoto}
\affiliation{%
  Department of Physics, Tokyo University of Science, Noda, Chiba 278-8510, Japan
}%


\date{\today}


\begin{abstract}
  We investigate Brownian motion with diffusivity alternately fluctuating
  between fast and slow states. We assume that sojourn-time distributions of
  these two states are given by exponential or power-law distributions. We
  develop a theory of alternating renewal processes to study a relaxation
  function which is expressed with an integral of the diffusivity over
  time. This relaxation function can be related to a position correlation
  function if the particle is in a harmonic potential, and to the
  self-intermediate scattering function if the potential force is absent.  It is
  theoretically shown that, at short times, the exponential relaxation or the
  stretched-exponential relaxation are observed depending on the power law index
  of the sojourn-time distributions. In contrast, at long times, a power law
  decay with an exponential cutoff is observed. The dependencies on the initial
  ensembles (i.e., equilibrium or non-equilibrium initial ensembles) are also
  elucidated. These theoretical results are consistent with numerical
  simulations.
\end{abstract}

\maketitle

\section {Introduction}
The Brownian motion, which is random motion of microscopic particles suspended
in fluids, is observed in various systems. The examples are colloid dynamics in
suspensions \cite{dhont96}, beads and center-of-mass motion of polymer chains
\cite{doi86}, and macromolecular diffusion in intracellular environments
\cite{bressloff13}.
The Brownian motion is usually characterized by mean square displacements
(MSDs). In fact, the MSD for a simple Brownian particle increases linearly with
time, and the proportionality factor is the diffusion coefficient multiplied by
$2d$ with $d$ being the spatial dimension of the system in which the Brownian
particle is immersed. The diffusion coefficient of a rigid particle depends on
its shape and the viscosity of the surrounding fluid
\cite{landau_fluid_mechanics, kim13}. For example, the diffusion coefficient of
a spherical particle with a small Reynolds number is given by the Stokes law, $D
= k_BT/6\pi\eta a$, with $k_{B}$ the Boltzmann constant, $T$ the absolute
temperature, $\eta$ the viscosity of the fluids, and $a$ the radius of the
particle.


The diffusion coefficient is usually considered as constant and is independent
of time. However, this is not always true; in crowded systems such as
supercooled liquids \cite{yamamoto98a, yamamoto98b} and intracellular
environments \cite{parry14}, the diffusion coefficient is not a constant, and
often behaves like a random variable. For such systems, the diffusion
coefficient should be interpreted as a time-dependent and fluctuating quantity.
The center-of-mass motion of polymer chains actually exhibits time-dependent and
fluctuating diffusivity \cite{uneyama12, uneyama15, miyaguchi17}. A prominent
feature of such Brownian motion with fluctuating diffusivity is that its
displacement distribution becomes non-Gaussian, though the MSD shows normal
diffusion (i.e., the linear time dependence of the MSD mentioned above)
\cite{mykyta14, uneyama15}.



Recently, Brownian motion with fluctuating diffusivity has been intensively
studied in the absence of the external force \cite{mykyta14, uneyama15, manzo15,
  chechkin17, miyaguchi17, jain17}, but it would be also important to
investigate the Brownian dynamics with some external forces, such as the forces
by some confinements and external potentials. For example, the diffusion
coefficient observed in single-particle-tracking experiments in bacterial
cytoplasms is typically of the order of $10^{-2}\,[\text{\textmu} \mathrm{m}^2/
\mathrm{sec}]$ \cite{golding06, parry14}, whereas the size of the bacterial cell
is of the order of $1 \text{\textmu} \mathrm{m}$. Thus, the time for diffusion
across a cell is of the order of $10\,\mathrm{sec}$ \cite{milo15}. On the other
hand, the measurement times of single-particle-tracking experiments are often
longer than $10\,\mathrm{sec}$ \cite{golding06, parry14}, and thus effects of
confinement inside the cells cannot be ignored.

In Ref.~\cite{uneyama19}, the authors analyzed the Brownian particle with
fluctuating diffusivity confined in a harmonic potential, that is, the
Ornstein-Uhlenbeck process with fluctuating diffusivity (OUFD). In particular, a
relaxation function $\Phi(t)$ of the position coordinates [See
Eq.~(\ref{e.def.relaxation-func}) below] was studied instead of the MSD [the MSD
is easily derived from $\Phi(t)$, if the system is in equilibrium]. In
Ref.~\cite{uneyama19}, the diffusion coefficient $D(t)$ is assumed to be
Markovian stochastic processes, and an eigenmode expansion for the relaxation
function $\Phi(t)$ is obtained through the eigenmode analysis of a transfer
operator. Formally, these results can be applicable to any systems with the
Markovian diffusivity $D(t)$. But it is also important to study systems with
non-Markovian diffusivity $D(t)$, because some properties (such as a large
scatter of a time-averaged MSD and weak ergodicity breaking) observed in
single-particle-tracking experiments \cite{golding06, weber10, weigel11, jeon11,
  burov11, tabei13, yamamoto14} are believed to originate from non-Markovian
properties of the system \cite{he08, meroz10, jeon11, burov11, tabei13,
  miyaguchi11b}.

In this paper, we study the Brownian motion with $D(t)$ being a non-Markovian
stochastic process.
Because general analysis for the non-Markovian diffusivity $D(t)$ is extremely
difficult, we assume that $D(t)$ takes only two values $D_+$ and $D_-$, and
randomly switches between these two states (the fast and slow states). It would
be worth noting that such two-state diffusion dynamics is actually a good
approximation to describe protein diffusion along DNA strands \cite{leith12},
large colloidal diffusion in bacterial cytoplasm \cite{parry14}, and movement
patterns of several kinds of bacteria \cite{detcheverry17}. To theoretically
analyze such two-state dynamics, we use the alternating renewal theory
\cite{cox62, goychuk03, akimoto15, miyaguchi16}, in which the system is assumed
to switch randomly between the two states. Similar theoretical framework for the
two-state dynamics is developed by utilizing non-commutable operator calculus in
Ref.~\cite{detcheverry17}.

Furthermore, sojourn-time distributions for the two states are assumed to be a
power law in this paper, which is a typical non-Markovian stochastic process.
As discussed in Ref.~\cite{miyaguchi16}, such power-law sojourn-time
distributions can originate from the cage effects in crowded systems
\cite{odagaki90, doliwa03, helfferich14a} or spatial heterogeneity.
Due to the power-law sojourn-time distribution, the relaxation function
$\Phi(t)$ shows exponential or stretched-exponential decay at short times,
whereas power-law decay with an exponential cutoff at long times.


This paper is organized as follows. In Sec.~\ref{s.model}, the OUFD and the
Langevin equation with fluctuating diffusivity (LEFD) are defined. In
Sec.~\ref{s.model}, the sojourn-time distribution and the initial ensembles are
also introduced. In Sec.~\ref{s.general-theory}, by utilizing an alternating
renewal theory, the relaxation function is represented with the sojourn-time
distributions. In Sec.~\ref{s.case-studies}, the relaxation function is derived
for the exponential and the power-law sojourn-time distributions. Finally,
Sec.~\ref{s.discussion} is devoted to a discussion. In Appendices, we summarize
some technical matters, including simulation details.

\section {Definition of the models}\label{s.model}

In this section, we define the OUFD, and introduce a position correlation
function $\Phi(t)$. In addition, the LEFD is also introduced, and it is shown
that the self-intermediate scattering function is given by a formula similar to
$\Phi(t)$.

\subsection {Ornstein-Uhlenbeck process with fluctuating diffusivity}

The Ornstein-Uhlenbeck process with fluctuating diffusivity is defined by the
following equation \cite{uneyama15, uneyama19}:
\begin{equation}
  \label{e.ou_proc.w.fluctuating-diffusivity}
  \frac {d \bm{r}(t)}{dt}
  =
  -\frac {\tilde{u} \bm{r}(t)}{\gamma(t)} + \sqrt{2D(t)} \bm{\xi}(t),
\end{equation}
where $\bm{r}(t)$ is the $d$-dimensional position vector of the Brownian
particle at time $t$, $\gamma(t)$ is a friction coefficient, and $\tilde{u}$ is
a constant which characterizes the strength of the restoring force [$\tilde{u}$
can be related to the spring constant of a harmonic potential $U(\bm{r})$, as
$U(\bm{r}) = \tilde{u} \bm{r}^2 / 2$].

The diffusion coefficient $D(t)$ is time dependent and fluctuating, i.e., $D(t)$
is a stochastic process. Moreover, $\bm{\xi}(t)$ is a white Gaussian noise which
satisfies%
\begin{equation}
  \label{e.white-gaussian-noise}
  \left\langle \bm{\xi}(t) \right\rangle_{\bm{\xi}} = 0, \qquad
  \left\langle \bm{\xi}(t) \bm{\xi}(t') \right\rangle_{\bm{\xi}}
  = \bm{I} \delta(t-t'),
\end{equation}
where $\bm{I}$ is the identity matrix, and we use the notation $\left\langle
\dots \right\rangle_{\bm{\xi}}$ to represent an ensemble average over the noise
history $\bm{\xi}(t)$. It is also assumed that $\bm{\xi}(t)$ and $D(t)$ are
mutually independent.

Because of Eq.~(\ref{e.white-gaussian-noise}), the thermal noise is
delta-correlated as
\begin{equation}
  \label{e.delta-correlated-thermal-noise}
  \left\langle
  \sqrt{D(t)} \bm{\xi}(t)
  \sqrt{D(t')} \bm{\xi}(t')
  \right\rangle_{\bm{\xi}}
  =
  D(t)
  \bm{I} \delta(t-t').
\end{equation}
We require that the system can reach the equilibrium state if the process $D(t)$
can reach the equilibrium state. This is equivalent to require that the
detailed-balance condition (or the local equilibrium condition) is satisfied for
the friction and noise coefficients \cite{sekimoto10}.
\begin{equation}
  \label{e.einstein-relation}
  \frac {k_BT}{\gamma(t)} = D(t).
\end{equation}

Multiplying both sides of Eq.~(\ref{e.ou_proc.w.fluctuating-diffusivity}) by
$\bm{r}(0)$, and averaging over a noise history $\bm{\xi}(t)$, we have an
ordinary differential equation
$d\left\langle \bm{r}(t) \cdot \bm{r}(0) \right\rangle_{\bm{\xi}}/dt = -u D(t)
\left\langle \bm{r}(t) \cdot \bm{r}(0) \right\rangle_{\bm{\xi}}$, where we
define $u$ as $u:= \tilde{u}/k_BT$ and used
$\left\langle \bm{\xi}(t)\cdot \bm{r}(0) \right\rangle_{\bm{\xi}} =
0$. Integrating this equation and averaging over realizations of $D(t)$ as well
as initial distributions of $\bm{r}(0)$, we have a relaxation function
\begin{equation}
  \label{e.def.relaxation-func}
  \Phi(t)
  :=
  \frac {
    \left\langle
    \bm{r}(t)\cdot \bm{r}(0)
    \right\rangle_{\bm{\xi},\mathrm{I}, D}
  }
  {\left\langle \bm{r}^2(0) \right\rangle_{\mathrm{I}}}
  =
  \left\langle
  \exp
  \left[
  -u \int_0^t dt' D(t')
  \right]
  \right\rangle_{D},
\end{equation}
where the subscripts $D$ and $\mathrm{I}$ represent the ensemble averages over
$D(t)$ and $\bm{r}(0)$, respectively.

It follows from Eq.~(\ref{e.def.relaxation-func}) that the relaxation function
$\Phi(t)$ is independent of the initial distributions of $\bm{r}(0)$; namely,
$\Phi(t)$ is characterized only by the stochastic process $D(t)$. In numerical
simulations, we use the local equilibrium distribution $(u/2\pi)^{d/2}
\exp\left[-u \bm{r}^2(0)/2\right]$ as an initial distribution of $\bm{r}(0)$,
but the results are independent of the choice of the initial distribution.
Therefore, whether $\bm{r}(0)$ follows the local equilibrium distribution or not
is unimportant concerning the relaxation function $\Phi(t)$. From
Eq.~(\ref{e.def.relaxation-func}), on the other hand, the relaxation function
clearly depends on $D(t)$ and thus the statistical properties of the stochastic
process $D(t)$ is important. In this work, we consider rather general stochastic
processes for $D(t)$ including the one that may not reach the equilibrium state
(See Sec.~\ref{s.case-studies.3}). We use the words \textit{equilibrium} and
\textit{non-equilibrium} to specify the initial ensembles of the stochastic
process $D(t)$. We will explain these initial ensembles in Sec.~\ref{s.model.3}
and Appendix \ref{s.equilibrium-ensemble}.

Moreover, if the system is in a stationary state---i.e., if $\bm{r}(0)$ follows
the local equilibrium distribution and $D(t)$ is a stationary stochastic
process---the MSD is given by \cite{uneyama19}
\begin{equation}
  \label{e.msd.relaxation-func}
  \left\langle | \bm{r}(t) - \bm{r}(0)|^2
  \right\rangle_{\bm{\xi}, \mathrm{I}, D}
  =
  2
  \left\langle \bm{r}^2(0) \right\rangle_{\mathrm{I}}
  \left[ 1 - \Phi(t) \right].
\end{equation}
Thus, the MSD is easily obtained from the relaxation function $\Phi(t)$, if the
system is stationary.

\subsection {Langevin equation with fluctuating diffusivity}

In this subsection, we show that the relaxation function $\Phi(t)$ given by
Eq.~(\ref{e.def.relaxation-func}) is related to the self-intermediate scattering
function for the LEFD
\begin{equation}
  \label{e.lengeven-eq.w.fluctuating-diffusivity}
  \frac {d \bm{r}(t)}{dt}
  =
  \sqrt{2D(t)} \bm{\xi}(t),
\end{equation}
where the potential term is absent, and thus it is translationally invariant in
space.  To study dynamical behavior of such a system, a density correlation
function $\left\langle \delta(\bm{r}'' - \bm{r}(t'')) \delta(\bm{r}' -
\bm{r}(t')) \right\rangle_{\bm{\xi}, D}$ is convenient. Due to the translational
symmetry, this correlation function depends on the position coordinates
$\bm{r}''$ and $\bm{r}'$ only through the displacement $\bm{r} := \bm{r}'' -
\bm{r}'$.  By also using a lag time $t := t'' - t'$, the correlation function is
rewritten as $\left\langle\delta(\bm{r} + \bm{r}' - \bm{r}(t +
t'))\delta(\bm{r}' - \bm{r}(t'))\right\rangle_{\bm{\xi}, D}$. Integrating with $
\bm{r}'$, we obtain \cite{hansen90}
\begin{equation}
  \label{e.van-Hove-correlation-func}
  G_s(\bm{r}, t)
  :=
  \left\langle
  \delta \left(
  \bm{r} - \left[ \bm{r}(t) - \bm{r}(0)\right]
  \right) 
  \right\rangle_{\bm{\xi}, \mathrm{I}, D},
\end{equation}
where we set $t'=0$ and take the average over $\bm{r}(0)$. Note that $D(t)$ is
not stationary in general, thus $G_s(\bm{r}, t)$ depends on the choice of the
origin of time. This distribution function for the particle displacement
$G_s(\bm{r}, t)$ is called the van Hove self-correlation function and it is
widely utilized to study the dynamical behavior of glass formers [From the
spherical symmetry, the angular integration of $G_s(\bm{r}, t)$ gives $4\pi
r^2G_s(r, t)$ with $r= |\bm{r}|$, and this latter function is usually utilized
\cite{kob95}].

Here, let us consider the same quantity with a given $D(t)$.
\begin{equation}
  \label{e.van-Hove-correlation-func.with-given-D(t)}
  G_s[\bm{r}, t | D]
  :=
  \left\langle
  \delta \left(
  \bm{r} - \left[ \bm{r}(t) - \bm{r}(0)\right]
  \right) 
  \right\rangle_{\bm{\xi}, \mathrm{I}}, 
\end{equation}
where the ensemble average for $D(t)$ is not taken.

Note that this correlation function is a functional of $D(t)$ and is related to
$G_s(\bm{r}, t)$ by
\begin{equation}
  \label{e.G_s(r,t)--G_s[r,t|D]}
  G_s(\bm{r}, t)
  =
  \left\langle G_s\!\left[\bm{r}, t | D\right] \right\rangle_{D}
  =
  \int \mathcal{D}\!D
  \,
  G_s\!\left[\bm{r}, t | D\right]
  \,
  \mathcal{P}\!\left[\mathcal{D}\right],
\end{equation}
where $\int \mathcal{D}\!D$ represents the path integral (functional integral)
over the diffusivity $D(t)$ and $\mathcal{P}[\mathcal{D}]$ is a path probability
of $D(t)$.

The functional $G_s[\bm{r}, t | D]$ is equivalent to a propagator of the
Brownian particle for a given $D(t)$, and thus it follows the Fokker-Planck
equation \cite{berne00}
\begin{equation}
  \label{e.fokker-planck-eq}
  \frac {\partial G_s[\bm{r}, t | D]}{\partial t}
  =
  D(t) \nabla^2 G_s[\bm{r}, t | D],
\end{equation}
with an initial condition $G_s[\bm{r}, 0 | D] = \delta(\bm{r})$. Here,
$\nabla^2$ is the Laplacian operator. This Fokker-Planck equation is solved by
using a Fourier transform
\begin{equation}
  \label{e.self-intermediate-scattering-func}
  F_s[\bm{k}, t|D]
  :=
  \int d\bm{r}
  e^{-i \bm{k}\cdot \bm{r}} G_s[\bm{r}, t | D].
\end{equation}
This functional is related to the self-intermediate scattering function as shown
below. The Fourier transform of Eq.~(\ref{e.fokker-planck-eq}) is given by
\begin{equation}
  \label{e.fokker-planck-eq.fourier}
  \frac {\partial F_s[\bm{k}, t | D]}{\partial t}
  =
  -D(t) k^2 F_s[\bm{k}, t | D],
\end{equation}
with an initial condition $F_s[\bm{k}, 0 | D] = 1$ and $k = |\bm{k}|$. This
differential equation is easily solved, and we obtain
\begin{equation}
  \label{e.Fs[k,t|D]}
  F_s[\bm{k}, t | D]
  =
  \exp\left[
  -k^2 \int_0^t D(t')dt'
  \right].
\end{equation}

The Fourier transform $F_s(\bm{k}, t)$ of the correlation function $G_s(\bm{r},
t)$ is usually referred to as the self-intermediate scattering function and
widely utilized to study dynamical behavior of glass formers. By using
Eq.~(\ref{e.Fs[k,t|D]}), the function $F_s(\bm{k}, t)$ is given by
\begin{equation}
    \label{e.Fs(k,t)}
    F_s(\bm{k}, t)
    =
    \left\langle 
    \exp\left[
    -k^2 \int_0^t D(t')dt'
    \right]
    \right\rangle_{D}.
  \end{equation}
Note that this function $F_s(\bm{k}, t)$ has the same form as $\Phi(t)$ in
Eq.~(\ref{e.def.relaxation-func}). A similar result was also derived in
Ref.~\cite{jain17} in a slightly different way.

\subsection {Definition of initial ensemble}\label{s.model.3}

In this subsection, the stochastic process $D(t)$ is defined as a two-state
process, and sojourn-time distributions for these states are introduced. If the
stochastic process $D(t)$ is a Markovian process, the system can reach the
equilibrium state if the detailed blance is satisfied (for the Markovian
processes, whether the detailed blance is satisfied or not can be checked rather
straightforwardly). On the other hand, in this work, we consider more general
processes for $D(t)$ which can be non-Markovian.  Thus the situation is (at
least apparently) not intuitive. Here we briefly explain the defintions of {\it
  equilibrium} and {\it non-equilibrium} initial states. The detailed
derivations for them are presented in Appendix~\ref{s.equilibrium-ensemble}.

\begin{figure}[t!]
  \centerline{\includegraphics[width=8.5cm]{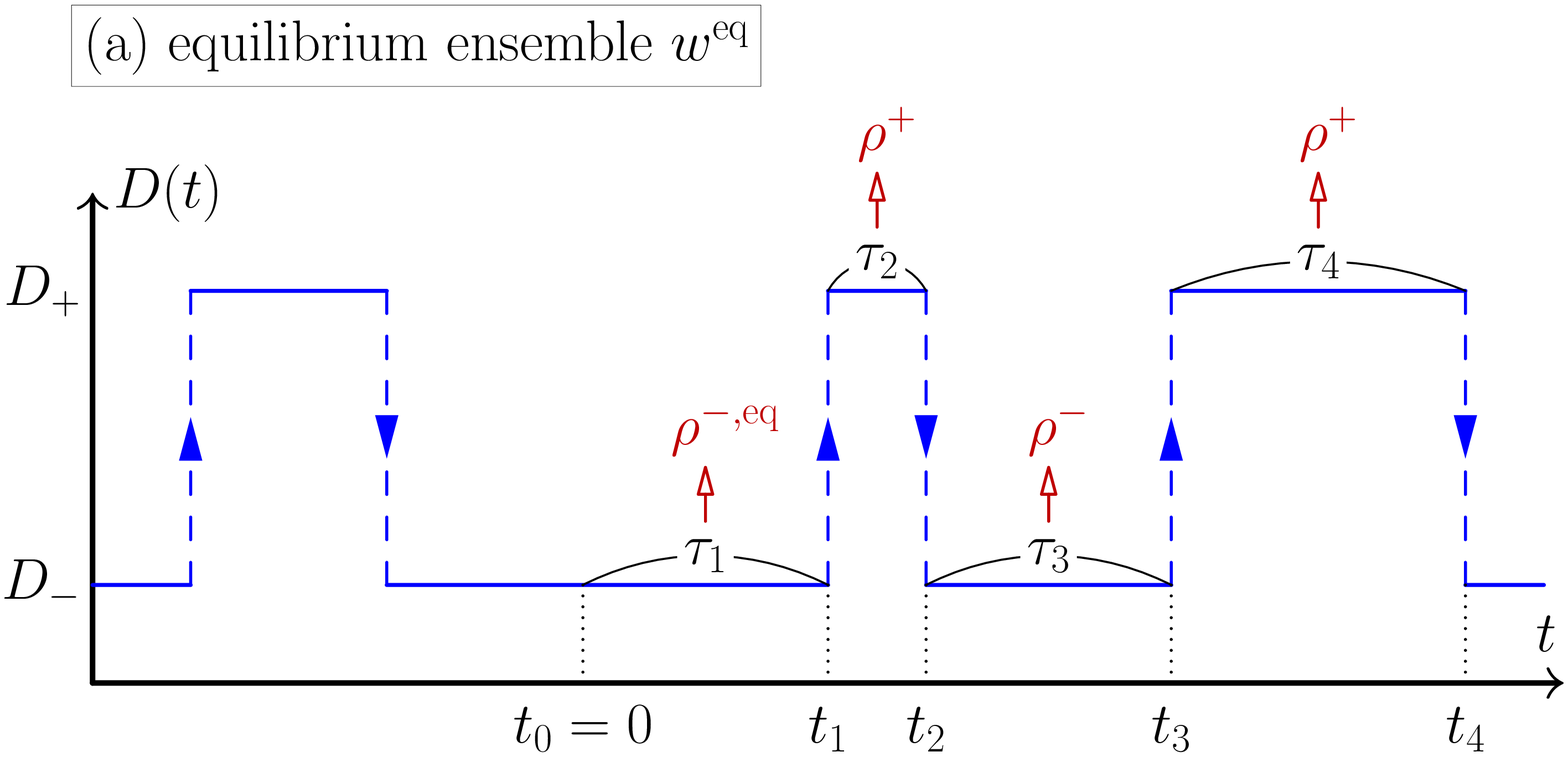}}
  \vspace*{.5cm}
  \centerline{\includegraphics[width=8.5cm]{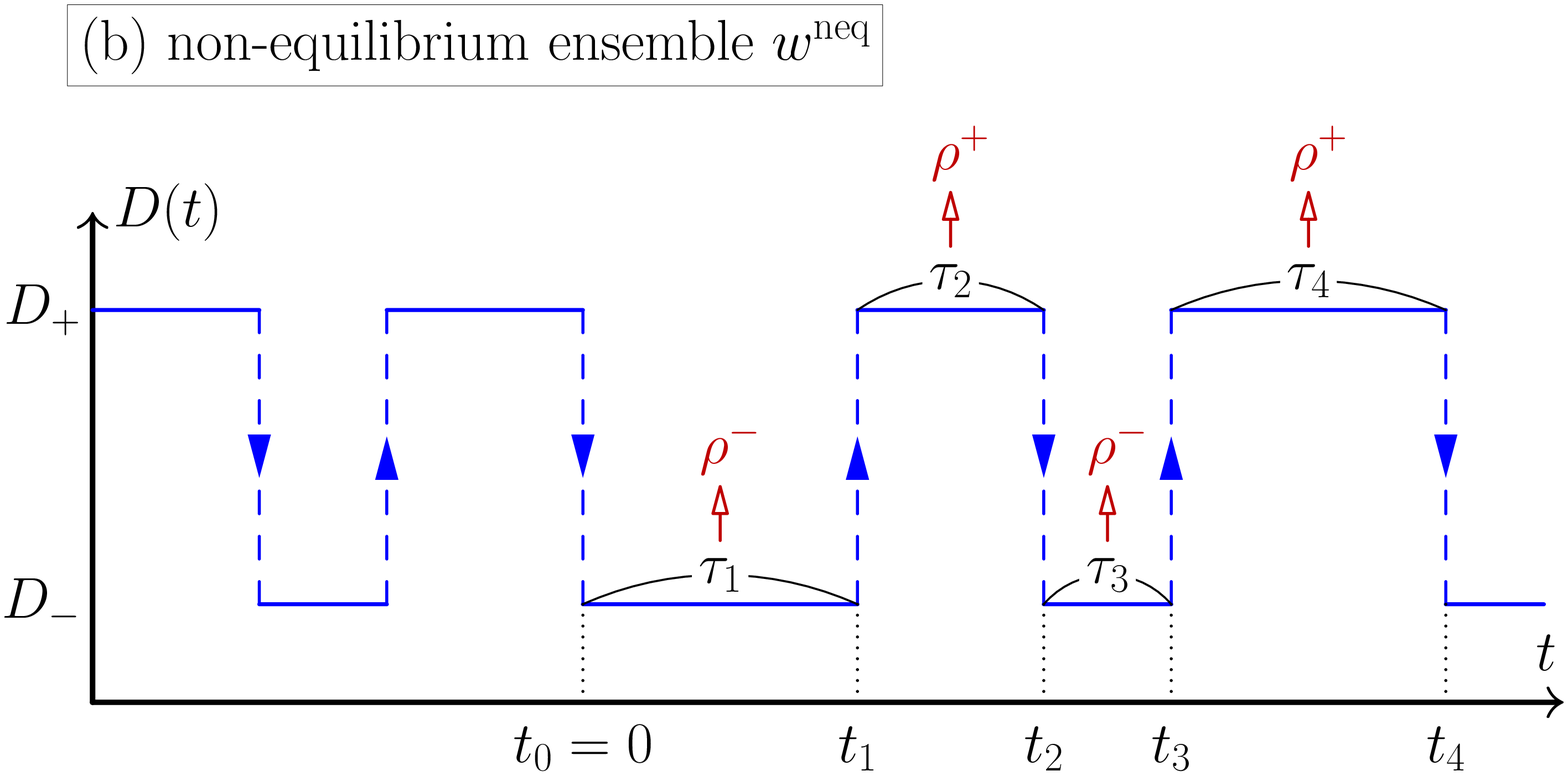}}
  \caption{\label{f.switching} (Color online) Schematic illustration of the
    switching process of diffusivity $D(t)$. (a) For the equilibrium ensemble
    $w^{\mathrm{eq}}(\tau_1)$, the process starts at $t=-\infty$, and thus there
    is no transition at $t=t_0$ in general. As an example, a realization with
    $D(0) = D_-$ is shown; For this trajectory, $\tau_{2n}\,(n=1,2,\dots)$
    follow the sojourn-time PDF $\rho^{+}(\tau)$, whereas
    $\tau_{2n+1}\,(n=1,2,\dots)$ follow $\rho^{-}(\tau)$. The first sojourn time
    $\tau_1$, however, follows the equilibrium PDF
    $\rho^{-, \mathrm{eq}}(\tau)$. (b) For the non-equilibrium ensemble
    $w^{\mathrm{neq}}(\tau_1)$, it is assumed that a transition occurs at
    $t=t_0$. As an example, a realization with $D(0) = D_-$ is shown; For this
    trajectory, $\tau_{2n}\,(n=1,2,\dots)$ follow the sojourn-time PDF
    $\rho^{+}(\tau)$, whereas $\tau_{2n+1}\,(n=0,1,\dots)$ follow
    $\rho^-(\tau)$.}
\end{figure}

In this paper, the diffusion coefficient $D(t)$ is assumed to be a dichotomous
process as illustrated in Fig.~\ref{f.switching}. Namely, at each time $t$, the
value of $D(t)$ is either of the two values $D_+$ or $D_-$:
\begin{equation}
  \label{e.def.D(t).two-state}
  D(t) :=
  \begin{cases}
    D_+ \quad (+ \,\text{state; fast state}),\\[.1cm]
    D_- \quad (- \,\text{state; slow sate}).
  \end{cases}
\end{equation}
In addition, we assume that $0 \leq D_- < D_+$. The condition $D_- \geq 0$ is
used for calculating double unilateral Laplace transforms in
Eq.~(\ref{e.f^R(u;s)-f^T+}) below. Contrastingly, in Ref.~\cite{godrche01}, a
similar calculation is carried out by using a bilateral Laplace transform,
thereby obtaining an equation corresponding to Eq.~(\ref{e.f^R(u;s)-f^T+}). This
difference in computation originates from the fact that, in
Ref.~\cite{godrche01}, a value corresponding to $D_-$ is set as $-1$ (In
Ref.~\cite{godrche01}, this corresponding variable is not the diffusion
coefficient, but it is related to the position of a diffusing particle. Thus,
the negative value is physically relevant).

We define the dichotomous process $D(t)$ by using the times at which the
transitions between the two states occur, $t_n\,(n=1, 2, \dots)$; these times
are referred to as renewal times [we also set $t_0=0$ for convenience, though
$t_0$ is not a renewal time in general. See Fig.~\ref{f.switching}(a)]. The
renewal time $t_n\,(n=1, 2, \dots)$ can be written by a sum of successive
sojourn times in the two states, $\tau_k$ ($k=1,2,\dots$), as
$t_n \equiv \sum_{k=1}^{n} \tau_k$.
A switching rule between the two states can then be given by sojourn-time
probability density functions (PDFs) of these states. Namely, the sojourn times
$\tau_k\,(k=2,3,\dots)$ for $+$ and $-$ states are random variables that follow
sojourn-time PDFs, $\rho^+(\tau)$ and $\rho^-(\tau)$, respectively.
%
%
%
%
To fully specify the process $D(t)$, we also need to define the PDF for the
first sojourn time $\tau_1$ (See Fig.~\ref{f.switching}). We mainly study two
kinds of the first sojourn-time PDFs: the equilibrium and non-equilibrium
ensembles.

\subsubsection {Equilibrium ensemble}

For the equilibrium ensemble, the process starts at $t=-\infty$, and thus the
system is in equilibrium at $t=0$. The diffusivity $D(0)$ is therefore at $+$ or
$-$ state with equilibrium fractions $p_{+}^{\mathrm{eq}}$ and
$p_{-}^{\mathrm{eq}}$. These fractions $p_{\pm}^{\mathrm{eq}}$ are simply given
by
\begin{equation}
  \label{e.eq-ensemble-1}
  p_{\pm}^{\mathrm{eq}}
  =
  \frac {\mu_{\pm}}{\mu_+ + \mu_-},
\end{equation}
where $\mu_{\pm}$ are mean sojourn times for $\rho^{\pm}(\tau)$. From this
expression, it is clear that the equilibrium ensemble exists only when the mean
sojourn times $\mu_{\pm}$ are finite; if $\mu_{\pm}$ diverges, the system cannot
reach the equilibrium state \cite{miyaguchi13}.

Moreover, the first sojourn time $\tau_1$ follows an equilibrium sojourn-time
PDF $\rho^{\pm, \mathrm{eq}}(\tau_1)$, not $\rho^{\pm}(\tau_{1})$. This is
because, in the equilibrium ensemble, $t=0$ is not a renewal time as illustrated
in Fig.~\ref{f.switching}(a).
Thus, the PDF of $\tau_1$ in the equilibrium ensemble is given by
\begin{equation}
  \label{e.def.eq.emsenble}
  w^{\mathrm{eq}}_{+}(\tau_1) + w^{\mathrm{eq}}_{-}(\tau_1)
  =
  p_{+}^{\mathrm{eq}}
  \rho^{+, \mathrm{eq}}(\tau_1) +
  p_{-}^{\mathrm{eq}}
  \rho^{-, \mathrm{eq}}(\tau_1),
\end{equation}
where $w^{\mathrm{eq}}_{\pm}(\tau_1)$ is defined by
$w^{\mathrm{eq}}_{\pm}(\tau_1):= p_{\pm}^{\mathrm{eq}} \rho^{\pm,
  \mathrm{eq}}(\tau_1)$. Namely, $\rho^{\pm, eq} (\tau_1)$ is a conditional PDF
of the first sojourn time $\tau_1$ given that the process starts at $t=-\infty$
and the state is $\pm$ at $t=0$. For a derivation and an explicit expression of
$\rho^{\pm, \mathrm{eq}}(\tau)$, see Appendix \ref{s.equilibrium-ensemble}.

\subsubsection {Non-equilibrium ensemble}

We also study a non-equilibrium ensemble. Here, we limit ourselves that a
transition occurs exactly at the inital time $t=0$ [See
Fig.~\ref{f.switching}(b)], and call such an ensemble as non-equilibrium. Note
that this is a typical non-equilibrium ensemble employed in the renewal theory
\cite{godrche01} and CTRW theory \cite{miyaguchi13}. In this ensemble, the
diffusivity $D(0)$ is at $+$ or $-$ state with some inital fractions $p_{+}^0$
and $p_{-}^0$; one of these fractions is arbitrary, and the other is given by
the normalization $p_+^0 + p_-^0 = 1$.

Moreover, if $D(0) = D_{\pm}$, the first sojourn time $\tau_1$ follows the
sojourn-time PDF $\rho^{\pm}(\tau_1)$; i.e., the first sojourn-time PDF is
the same as those for the following sojourn times $\tau_n\,(n=3, 5, \dots)$.
This is because we assume that $t=0$ is a renewal time for the non-equilibrium
ensemble. It follows that, the first sojourn-time PDF is given by
\begin{equation}
  \label{e.def.neq.emsenble}
  w^{\mathrm{neq}}_{+}(\tau_1) + w^{\mathrm{neq}}_{-}(\tau_1)
  =
  p_{+}^0 \rho^{+}(\tau_1) + p_{-}^0 \rho^{-}(\tau_1),
\end{equation}
where $w^{\mathrm{neq}}_{\pm}(\tau_1)$ is defined by
$w^{\mathrm{neq}}_{\pm}(\tau_1) := p_{\pm}^0 \rho^{\pm}(\tau_1)$.

\subsubsection {General ensemble}

In the next section, we present a theory for a more general inital ensemble
rather than the specific ensembles defined above. Let us denote this general
inital ensemble as
\begin{equation}
  \label{e.def.genelal.emsenble}
  w^0_{+}(\tau_1) + w^0_{-}(\tau_1)
  =
  p_{+}^0 \rho^{+,0}(\tau_1) + p_{-}^0 \rho^{-,0}(\tau_1),
\end{equation}
where $w^0_{\pm}(\tau_1):= p_{\pm}^0 \rho^{\pm, 0}(\tau_1)$, and
$\rho^{\pm,0}(\tau)$ is the first sojourn-time PDF given that the initial state
is $\pm$. Therefore, if we replace $p_{\pm}^{0}$ and $\rho^{\pm,0}(\tau_1)$ in
Eq.~(\ref{e.def.genelal.emsenble}) with $p_{\pm}^{\mathrm{eq}}$ and $\rho^{\pm,
  \mathrm{eq}}(\tau_1)$, we obtain the equilibrium ensemble defined in
Eq.~(\ref{e.def.eq.emsenble}); similarly, if we replace $\rho^{\pm,0}(\tau_1)$
in Eq.~(\ref{e.def.genelal.emsenble}) with $\rho^{\pm}(\tau_1)$, we obtain the
non-equilibrium ensemble defined in Eq.~(\ref{e.def.neq.emsenble}). These
replacement rules are used in the following sections to obtain results for the
equilibrium and non-equilibrium ensembles.

We also use notations such as $w^0(\tau_1) := w_+^{0}(\tau_1) +
w_-^{0}(\tau_1)$, $w^{\mathrm{eq}}(\tau_1) := w_+^{\mathrm{eq}}(\tau_1) +
w_-^{\mathrm{eq}}(\tau_1)$ and $w^{\mathrm{neq}}(\tau_1) :=
w_+^{\mathrm{neq}}(\tau_1) + w_-^{\mathrm{neq}}(\tau_1)$, with which we can
completely specify the initial ensemble for $D(t)$. Moreover, instead of the
notation $\left\langle \dots \right\rangle_D$ used in the previous subsections,
hereafter we employ the following notations for the ensemble average over
$D(t)$. We denote the average in terms of the equilibrium ensembles
$w^{\mathrm{eq}}(\tau_1)$ as $\left\langle \dots \right\rangle_{\mathrm{eq}}$,
and the average in terms of the genelal initial ensemble $w^0(\tau_1)$ as
$\left\langle \dots \right\rangle$ (the bracket without a subscript).

\section {Alternating renewal theory: general analysis}\label{s.general-theory}

In this section, we derive a general formula for the relaxation function
$\Phi(t)$ by using the alternating renewal theory \cite{cox62, miyaguchi16}, in
which the system has two states with different sojourn-time PDFs in general. In
fact, we show that $\Phi(t)$ is closely related to a quantity which is called
occupation times in the renewal theory \cite{godrche01}.

\subsection {Definition of occupation time}

To explicitly represent the variables and initial conditions, let us denote the
relaxation function $\Phi(t)$ as $\hat{f}^D(u; t | w^0)$; namely [see
Eq.~(\ref{e.def.relaxation-func})]
\begin{equation}
  \label{e.def.f^R(u;t)}
  \hat{f}^D(u; t | w^0) :=
  \left\langle
  \exp
  \left[
  -u \int_0^t dt' D(t')
  \right]
  \right\rangle.
\end{equation}
Also, we define an integral quantity $\bar{D}(t)$ as
\begin{align}
  \bar{D}(t)
  := \int_{0}^{t} D(t')dt' 
  & = D_+T_+(t) + D_-T_-(t) \notag \\[0.1cm]
  \label{e.R(t)-T+(t)}
  & = \left(D_{+} - D_{-}\right)T_+(t) + D_-t,
\end{align}
where the occupation times $T_+(t)$ and $T_-(t)$ are the total times spent by
the system in the $+$ and $-$ states up to time $t$, respectively.
Then, $\hat{f}^D(u; t | w^0) = \left\langle \exp(- u \bar{D}(t)) \right\rangle$
is the Laplace transform of the PDF $f^D(\bar{d}; t | w^0)$ with the initial
ensemble $w^0 (\tau_1)$. Note that if $D_+=1$ and $D_- = -1$, $\bar{D}(t)$ is
often referred to as a mean magnetization by analogy between the two-state
trajectory $D(t) = \pm 1$ and a spin configuration \cite{godrche01}.

Let us denote the PDF of $T_{\pm}(t)$ with an initial ensemble $w^0 (\tau_1)$ as
$f^{T_{\pm}} (t_{\pm}; t | w^0)$. Here, we focus on the PDF $f^{T_+} (t_+; t |
w_0)$, because $f^{T_-} (t_{-}; t | w^0)$ can be readily obtained from $f^{T_+}
(t_{+}; t | w_0)$ by
\begin{equation}
  f^{T_-} (t_-; t | w^0)
  =
  f^{T_+} (t-t_-; t | w^0).
\end{equation}
Similarly, from Eq.~(\ref{e.R(t)-T+(t)}), the PDF $f^D(\bar{d}; t | w^0)$ is
also given by $f^{T_+} (t_+; t | w^0)$ as
\begin{equation}
  \label{e.f^R(r;t)-f^T+}
  f^D(\bar{d}; t | w^0)
  =
  \frac {1}{{D_+-D_-}}
  f^{T_+} \left(\left.\frac {\bar{d}-D_{-}t}{D_+-D_-}; t \right|w^0\right)
\end{equation}
Carrying out double Laplace transforms of Eq.~(\ref{e.f^R(r;t)-f^T+}) in terms
of $\bar{d} \leftrightarrow u$ and then $t \leftrightarrow s$, we have
\begin{equation}
  \label{e.f^R(u;s)-f^T+}
  \breve{f}^D(u;s| w^0)
  =
  \breve{f}^{T_+} \left( (D_+-D_-)u; s + D_-u | w^0\right),
\end{equation}
where we used the assumption $D_-\geq 0$ in the Laplace transform with respect
to $\bar{d}$.

\subsection {PDF of occupation time}

To derive an explicit form of $\breve{f}^{T_+} \left(u; s | w^0\right)$, we
define $f^{T_+}_{\pm,n} (t_+; t | w^0)$ as a conditional joint PDF that (i) the
state is $\pm$ at time $t=0$, (ii) the number of the renewals up to time $t$ is
$n$, and (iii) the occupation time of the $+$ state up to time $t$ is $t_+$
under the condition that the initial ensemble is $w_0$.
The function $\breve{f}^{T_{+}} (u; s | w^0)$ can be expressed by using
$\breve{f}^{T_+}_{\pm, n} (u; s | w^0)$ as
\begin{align}
  \breve{f}^{T_{+}} (u; s | w^0)
  =
  \sum_{n=0}^{\infty}
  &
  \left[
  \breve{f}^{T_{+}}_{+,2n} (u; s | w^0 )
  +
  \breve{f}^{T_{+}}_{+,2n+1} (u; s | w^0)
  \right.
  \notag\\[0.1cm]
  \label{e.f^T+(u;s).sum}
  +&
  \breve{f}^{T_{+}}_{-,2n}(u; s | w^0 )
  +
  \left.
  \breve{f}^{T_{+}}_{-,2n+1} (u; s | w^0)
  \right].
\end{align}

Now, we derive each term in the right-hand side of Eq.~(\ref{e.f^T+(u;s).sum}).
The joint PDF $f^{T_+}_{\pm, n} (t_+; t | w^0)$ can be written as
\cite{godrche01}
\begin{equation}
  \label{e.f^T+(t+;t)}
  f^{T_+}_{\pm, n} (t_+; t | w^0)
  =
  \left\langle
  \delta_{\pm}
  \delta\left( t_+ - T_+(t) \right)
  I(t_n \leq t < t_{n+1})
  \right\rangle,
\end{equation}
where $I(\dots)$ and $\delta_{\pm}$ are quantities that take $1$ or $0$.
$I(\dots)$ becomes $1$ if the inside of the bracket is satisfied, and otherwise
becomes 0. Moreover, $\delta_{\pm} := \delta_{D(0),D_{\pm}}$ is a random
variable indicating the initial state, i.e.,
\begin{equation}
  \delta_{\pm} =
  \begin{cases}
    1, \quad\, & \text{if the state is $\pm$ at $t=0$},\\
    0, \quad\, & \text{if the state is $\mp$ at $t=0$}.
  \end{cases}
\end{equation}
By taking the Laplace transforms of Eq.~(\ref{e.f^T+(t+;t)}) in terms of $t_+
\leftrightarrow u$ and $t \leftrightarrow s$, we have
\begin{align}
  \breve{f}^{T_+}_{\pm, n} (u; s | w^0)
  &=
  \left\langle
  \delta_{\pm}
  \right\rangle
  \left\langle
  \int_{t_n}^{t_{n+1}} dt e^{-uT_+(t)} e^{-st}
  \right\rangle_{\pm}
  \notag\\[0.1cm]
  \label{e.f^T+(u;s)}
  &=
  p_{\pm}^0
  \left\langle
  \int_{t_n}^{t_{n+1}} dt e^{-uT_+(t)} e^{-st}
  \right\rangle_{\pm},
\end{align}
where $\left\langle \dots \right\rangle_{\pm}$ is a conditional average under
the condition that the initial state is $\pm$.

The ensemble average in Eq.~(\ref{e.f^T+(u;s)}) can be calculated as follows.
If the initial state is the $+$ state [i.e., $D(0)= D_+$], the occupation time
$T_+(t)$ can be expressed as
\begin{equation}
  T_+(t) = \tau_1 + \tau_3 + \dots + \tau_{2n-1} + (t-t_{2n}),
\end{equation}
when $2n$ transitions occur up to time $t$, or
\begin{equation}
  T_+(t) = \tau_1 + \tau_3 + \dots + \tau_{2n-1},
\end{equation}
when $2n-1$ transitions occur up to time $t$. Thus, from
Eq.~(\ref{e.f^T+(u;s)}), for $n=0$ we have 
\begin{equation}
  \label{e.f^T+_+0}
  \breve{f}^{T_+}_{+, 0} (u; s | w^0)
  =
  \frac {p_+^0 - \hat{w}_+^0(u+s)}{u+s},
\end{equation}
where we used $p_+^0\left\langle e^{-(u+s)\tau_1}\right\rangle =
p_+^0\hat{\rho}^{+,0}(u+s) = \hat{w}_+^0(u+s)$. For $n=1,2,\dots$, we can obtain
$\breve{f}^{T_+}_{+, n} (u; s | w^0)$ with a calculation similar to
Eq.~(\ref{e.app.<delta.exp>_2n-1}) in Appendix \ref{s.equilibrium-ensemble}.
From Eq.~(\ref{e.f^T+(u;s)}), we have
\begin{align}
  \label{e.f^T+_+2n}
  &\breve{f}^{T_+}_{+, 2n} (u; s | w^0)
  =
  \hat{w}_+^0(u+s)
  \hat{\rho}^{n-1}(u,s)
  \hat{\rho}^{-}(s)
  \frac {1-\hat{\rho}^{+}(u+s)}{u+s},
  \\[0.1cm]
  \label{e.f^T+_+2n-1}
  &\breve{f}^{T_+}_{+, 2n-1} (u; s | w^0)
  =
  \hat{w}_+^0(u+s)
  \hat{\rho}^{n-1}(u,s)
  \frac {1-\hat{\rho}^{-}(s)}{s},
\end{align}
for $n=1,2,\dots$. Here, the function $\hat{\rho} (u, s)$ is
defined as $\hat{\rho} (u, s):= \hat{\rho}^{+}(u+s)\hat{\rho}^{-}(s)$.

Similarly, if the initial state is the $-$ state [i.e., $D(0)= D_-$], we have
\begin{equation}
  T_+(t) = \tau_2 + \tau_4 + \dots + \tau_{2n},
\end{equation}
when $2n$ transitions occur up to time $t$, or
\begin{equation}
  T_+(t) = \tau_2 + \tau_4 + \dots + \tau_{2n-2} + (t-t_{2n-1}),
\end{equation}
when $2n-1$ transitions occur up to time $t$. Thus, from
Eq.~(\ref{e.f^T+(u;s)}), we have
\begin{align}
  \label{e.f^T+_-0}
  &\breve{f}^{T_+}_{-, 0} (u; s | w^0)
  =
  \frac {p_-^0 - \hat{w}_-^0(s)}{s},
  \\[0.1cm]
  \label{e.f^T+_-2n}
  &\breve{f}^{T_+}_{-, 2n} (u; s | w^0)
  =
  \hat{w}_-^0(s)
  \hat{\rho}^{n-1}(u,s)
  \hat{\rho}^{+}(u+s)
  \frac {1-\hat{\rho}^{-}(s)}{s},
  \\[0.1cm]
  \label{e.f^T+_-2n-1}
  &\breve{f}^{T_+}_{-, 2n-1} (u; s | w^0)
  =
  \hat{w}_-^0(s)
  \hat{\rho}^{n-1}(u,s)
  \frac {1-\hat{\rho}^{+}(u+s)}{u+s},
\end{align}
for $n=1,2,\dots$.

Substituting Eqs.~(\ref{e.f^T+_+0})--(\ref{e.f^T+_+2n-1}) and
Eqs.~(\ref{e.f^T+_-0})--(\ref{e.f^T+_-2n-1}) into (\ref{e.f^T+(u;s).sum}), we
have
\begin{align}
  \breve{f}^{T_{+}} (u; s | w^0)
  =&
  \frac {p_+^0}{u+s}
  +
  \frac
  {\left(\frac {1}{s} - \frac {1}{u+s}\right)\hat{w}_+^0(u+s) \left[1 - \hat{\rho}^-(s)\right]}
  {1- \hat{\rho}^+(u+s) \hat{\rho}^-(s)}
  \notag\\[0.1cm]
  \label{e.f^T+(u;s).general}
  +&\frac {p_-^0}{s}
  +
  \frac
  {\left(\frac {1}{u+s} - \frac {1}{s}\right)\hat{w}_-^0(s)\left[1 - \hat{\rho}^+(u+s)\right]}
  {1- \hat{\rho}^+(u+s) \hat{\rho}^-(s)}.
\end{align}
Now we can express the PDF of the occupation time $T^+(t)$ by using the initial
fractions $p_{\pm}^0$, and the first and subsequent sojourn-time PDFs,
$\hat{w}^0_{\pm}(s)$ and $\hat{\rho}^{\pm}(s)$.

\subsection {Explicit form of relaxation function}

Finally, we derive explicit forms of the relaxation function. From
Eqs.~(\ref{e.f^R(u;s)-f^T+}) and (\ref{e.f^T+(u;s).general}), we have
\begin{align}
  \label{e.f^R(u;s).general}
  \breve{f}^{D} (u; s | w^0)
  =&
  \frac {p_+^0}{z_+}
  +
  \frac
  {\left(\frac {1}{z_-} - \frac {1}{z_+}\right)\hat{w}_+^0(z_+) \left[1 - \hat{\rho}^-(z_-)\right]}
  {1- \hat{\rho}^+(z_+) \hat{\rho}^-(z_-)}
  \notag\\[0.1cm]
  +&\frac {p_-^0}{z_-}
  +
  \frac
  {\left(\frac {1}{z_+} - \frac {1}{z_-}\right)\hat{w}_-^0(z_-)\left[1 - \hat{\rho}^+(z_+)\right]}
  {1- \hat{\rho}^+(z_+) \hat{\rho}^-(z_-)},
\end{align}
where $z_{\pm}$ is defined by $z_{\pm} := D_{\pm}u + s$.
Here, let us define functions $\tilde{\rho}^{\pm}(s)$ as
\begin{equation}
  \label{e.tilde.rho}
  \tilde{\rho}^{\pm}(s) := 1 -   \hat{\rho}^{\pm}(s).
\end{equation}
By using these auxiliary functions, Eq.~(\ref{e.f^R(u;s).general}) can be
rewritten as
\begin{align}
  \label{e.f^R(u;s).general.2}
  \breve{f}^{D} (u; s | w^0)
  =&
  \frac {p_+^0}{z_+}
  +
  \frac {p_+^0\left(\frac {1}{z_-} - \frac {1}{z_+}\right)\hat{\rho}^{+,0}(z_+) \tilde{\rho}^-(z_-)}
  {\tilde{\rho}^+(z_+) + \tilde{\rho}^-(z_-) - \tilde{\rho}^+(z_+) \tilde{\rho}^-(z_-)}
  \notag\\[0.1cm]
  +&
  \frac {p_-^0}{z_-}
  +
  \frac {p_-^0\left(\frac {1}{z_+} - \frac {1}{z_-}\right)\hat{\rho}^{-,0}(z_-) \tilde{\rho}^+(z_+)}
  {\tilde{\rho}^+(z_+) + \tilde{\rho}^-(z_-) - \tilde{\rho}^+(z_+) \tilde{\rho}^-(z_-)},
\end{align}
where we used $\hat{w}_{\pm}^0(s) = p_{\pm}^0 \hat{\rho}^{0,\pm}(s)$.  If the
two diffusion coefficients are the same $D_+ = D_- =: D$, we have $z_+ = z_-$;
and thus the second and fourth of terms in Eq.~(\ref{e.f^R(u;s).general.2})
vanish. The first and third terms give $\breve{f}^{D} (u; s | w^0) = 1/(s+Du)$,
and thus we obtain an exponential decay: $\hat{f}^{D} (u; t | w^0) =
e^{-Dut}$. This result can be also obtained more directly from
Eq.~(\ref{e.def.f^R(u;t)}).

The double Laplace transform $\breve{f}^D (u; s | w^{\mathrm{eq}})$ for the
equilibrium ensemble $w^{\mathrm{eq}}$ can be obtained simply by
replacing $\hat{\rho}^{\pm, 0}$ and $p_{\pm}^0$ with $\hat{\rho}^{\pm,
  \mathrm{eq}}$ and $p_{\pm}^{\mathrm{eq}}$ in Eq.~(\ref{e.f^R(u;s).general.2}),
respectively. Similarly, $\breve{f}^D (u; s | w^{\mathrm{neq}})$ for the
non-equilibrium ensemble is obtained by replacing $\hat{\rho}^{\pm, 0}$ with
$\hat{\rho}^{\pm}$.


\section {Case studies}\label{s.case-studies}

In this section, we study the relaxation function $\hat{f}^D(u; t | w^0)$ for
the cases in which the sojourn-time PDFs are given by the exponential
distributions or the power-law distributions.  For the case where the mean
sojourn times exist (i.e., $\mu_{\pm} < \infty$), we assume that the switching
between the two states is fast compared with the diffusive time scale---i.e.,
$\mu_{\pm} \ll {1}/{D_+u}$. Moreover, we focus on the time regime
$t \gtrsim {1}/{D_+u}$; in the Laplace domain, this corresponds to
$s \lesssim D_+u$. These fast switching assumptions can be rewritten as
\begin{equation*}
  \mu_{\pm} \ll \frac {1}{D_+u} \sim \frac {1}{D_+u + s} = \frac {1}{z_+} < \frac {1}{z_-}.  
\end{equation*}
Thus, we obtain $\mu_{\pm} \ll 1/z_{\pm}$ or equivalently $z_{\pm} \ll k_{\pm}$,
where $k_{\pm}$ is the transition rate and given by $k_{\pm} = 1/\mu_{\pm}$.

\subsection {Exponential distribution}

If both sojourn-time PDFs, $\rho^{\pm}(\tau)$, are given by the exponential
distributions, the first sojourn-time PDF is given by
\begin{equation}
  \label{e.w0.exp-dist.non-equilibrium}
  w^0(\tau_1)
  = p^0_+ \rho^{+,0}(\tau_1) + p^0_-\rho^{-,0}(\tau_1)
  =
  p^0_+ \rho^+(\tau_1) + p^0_-\rho^-(\tau_1),
\end{equation}
Therefore, in this case, the general initial ensemble $w^0 (\tau_1)$ is
equivalent to the typical non-equilibrium ensemble $w^{\text{neq}}(\tau_1)$.
This equivalence is due to the fact that the exponential distributions are
Markovian; a direct proof is possible as follows. Generally,
$\rho^{\pm,0}(\tau_1)$ should be given by a forward-recurrence-time PDF with
respect to $\rho^{\pm}(\tau)$, but this forward-recurrence-time PDF can be shown
to be equivalent to $\rho^{\pm}(\tau)$ if $\rho^{\pm}(\tau)$ is an exponential
distribution due to its Markovian property [See
Eq.~(\ref{e.forward-reccurence-PDF.exponential.w(tau;t')}) in Appendix
\ref{s.equilibrium-ensemble}].
By using Eqs.~(\ref{e.w0.exp-dist.non-equilibrium}), (\ref{e.laplace.exp-dist}),
and (\ref{e.laplace.exp-dist.stationality}), $\breve{f}^D (u; s | w^0)$
in Eq.~(\ref{e.f^R(u;s).general}) can be rewritten as
\begin{equation}
  \label{e.f^R(u;s).exp-dist.general}
  \breve{f}^{D} (u; s | w^0)
  =
  \frac
  {p_+^0 (k_+ + k_- + z_-) + p_-^0 (k_+ + k_- + z_+)}
  {(k_+ + z_+)(k_- + z_-) - k_+k_-}.
\end{equation}
This function is equivalent to that for the typical non-equilibrium ensemble
$\breve{f}^{D} (u; s | w^{\text{neq}})$.

For the case of the equilibrium ensemble, we have $ w^{\mathrm{eq}}(\tau_1) =
p^{\mathrm{eq}}_+ \rho^{+}(\tau_1) + p^{\mathrm{eq}}_- \rho^{-}(\tau_1)$, and
Eq.~(\ref{e.f^R(u;s).exp-dist.general}) can be further simplified by replacing
$p_{\pm}^0$ with $p_{\pm}^{\mathrm{eq}}$ as
\begin{equation}
  \label{e.f^R(u;s).exp-dist.equilibrium}
  \breve{f}^{D} (u; s | w^{\mathrm{eq}})
  =
  \frac
  {k_+ z_+ + k_- z_- + (k_+ + k_-)^2}
  {\left(k_+ + k_-\right)
    \left[(k_+ + z_+)(k_- + z_-) - k_+k_-\right]}.
\end{equation}
This equation is consistent with the result obtained in Ref.~\cite{uneyama19},
in which a transfer operator method is utilized. In fact, the relaxation
function $\breve{f}^{D} (u; s | w^{\mathrm{eq}})$ has two simple poles in
terms of $s$, which correspond to two different relaxation modes
\cite{uneyama19}.

If we assume that the switching between the two-state is fast, i.e., $z_{\pm}
\ll k_{\pm}$, we have
\begin{align}
  \label{e.f^R(u;s).exp-dist.equilibrium.fast-switching}
  \breve{f}^{D} (u; s | w^{\mathrm{eq}})
  &\simeq
  \frac
  {k_+ + k_-}{k_+ z_- + k_- z_+}
  \notag\\[0.1cm]
  &=
  \frac {1}{\left\langle D \right\rangle_{\mathrm{eq}} u + s}
  \quad
  \underset{
    \begin{subarray}{c}
      \mathcal{L}^{-1}\\
      (s \rightarrow t)
    \end{subarray}}{\longrightarrow}
  \quad
  e^{-\left\langle D \right\rangle_{\mathrm{eq}} u t},
\end{align}
where $\left\langle D \right\rangle_{\mathrm{eq}}$ is defined by $\left\langle D
\right\rangle_{\mathrm{eq}} := D_+ p_+^{\mathrm{eq}} + D_- p_-^{\mathrm{eq}}$.
Therefore, in this limit, we obtain a single mode relaxation, as if the particle
diffuses with the diffusion coefficient $\left\langle D
\right\rangle_{\mathrm{eq}}$.

The same result is obtained from the relaxation function with the
non-equilibrium initial ensemble [Eq.~(\ref{e.f^R(u;s).exp-dist.general})]. This
is because, under the assumption of the fast switching $\left\langle
D\right\rangle_{\mathrm{eq}} u \ll k_{\pm} = 1/\mu_{\pm}$, the relaxation of the
fractions [Eq.~(\ref{e.forward-reccurence-PDF.exponential.p(t')})] is much
faster than that given in
Eq.~(\ref{e.f^R(u;s).exp-dist.equilibrium.fast-switching}).

\subsection {Power-law distribution: $1<\alpha_-<2$}

Let us study the case where the sojourn-time PDFs are given by the power-law
distributions $\rho^{\pm}(\tau) \sim 1/\tau^{\alpha_{\pm}}$ with $1<
\alpha_{\pm}\leq 2$. In this case, the mean sojourn times exist, and thus the
equilibrium ensemble can be defined. In this and the next subsections, we assume
that $D_- \ll D_+$.


\begin{figure*}[t!]
  \begin{minipage}[bt]{8.5cm}
    \centerline{\includegraphics[width=8.0cm]{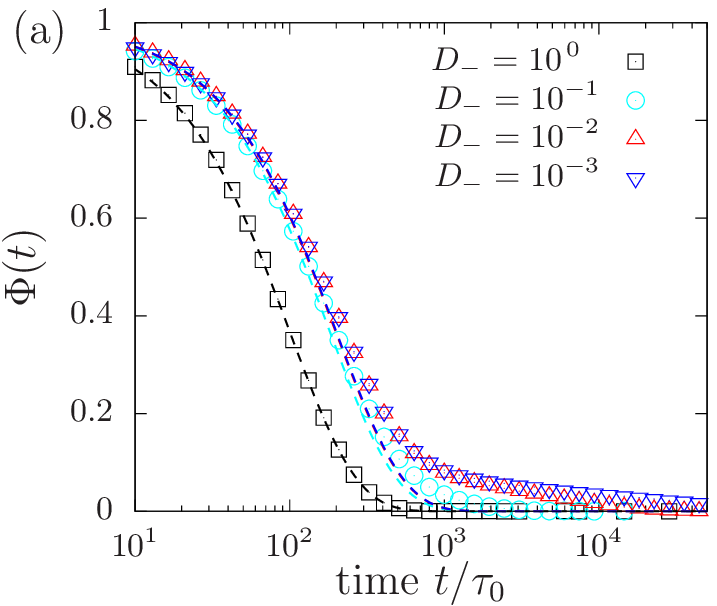}}
  \end{minipage}
  \begin{minipage}[bt]{8.5cm}
    \centerline{\includegraphics[width=8.0cm]{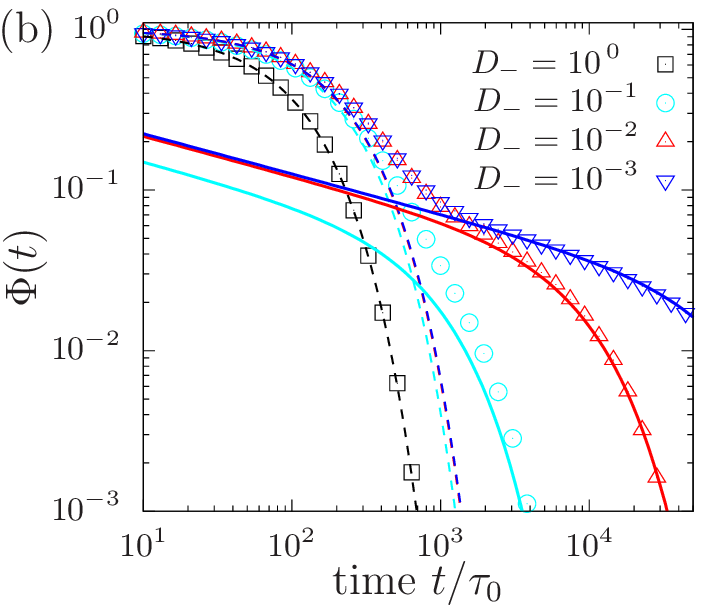}}
  \end{minipage}
  \caption{\label{f.power-law(1,2).eq} (Color online) Relaxation function
    $\Phi(t)$ as a function of time $t$ in (a) semi-log form and (b) log-log
    form. Time is measured in units of $\tau_0$, a cutoff time for short trap
    times (See Appendix \ref{sec:app.simulation} for details). The values of
    $\Phi(t)$ are obtained by numerical simulations of
    Eq.~(\ref{e.ou_proc.w.fluctuating-diffusivity}) with the spatial dimension
    $d=1$. The initial ensemble for $D(t)$ is assumed to be in equilibrium, and
    the initial distribution for $r(t)$ is given by the local equilibrium
    ensemble. Results for four different values of the slow diffusion
    coefficient $D_-$ are displayed: $D_-/D_+ = 10^0 \,(\text{square}),
    10^{-1}\,(\text{circle}), 10^{-2}\,(\text{triangle up}),$ and $10^{-3}
    \,(\text{triangle down})$. The damping coefficient $u$ is set as $u =
    0.01/D_+\tau_0$. The scaling exponents $\alpha_{-}$ of the power law
    distribution $\rho^{-}(\tau)$ is set as $\alpha_-= 1.25$. In contrast,
    $\rho^{+}(\tau)$ is assumed to be the exponential distribution with $\mu_+ =
    \mu_-$. The dashed and solid lines are the theoretical predictions for the
    short and long time regimes, respectively
    [Eq.~(\ref{e.f^R(u;t).power-law(1,2).eq.short}) and
    Eq.~(\ref{e.f^R(u;t).power-law(1,2).eq.long})]. There are no fitting
    parameters.}
\end{figure*}

\subsubsection {Equilibrium ensemble}
If the sojourn-time distributions $\rho^{\pm}(\tau)$ follow the power-law
distributions with finite means $\mu_{\pm}$, their Laplace transforms ($\tau
\leftrightarrow \sigma$) can be expressed as
\begin{equation}
  \label{e.rho(s).power-law(1,2)}
  \hat{\rho}^{\pm}(\sigma)
  \underset{\sigma \to 0}{\simeq}  1
  - \mu_{\pm} \sigma + a_{\pm} \sigma^{\alpha_{\pm}}  + o(\sigma^{\alpha_{\pm}}),
\end{equation}
where $1< \alpha_- < \alpha_+ \leq 2$, and $o(\sigma^{\alpha_{\pm}})$ is
Landau's notation. When $\alpha_+ = 2$, $\rho^+(\tau)$ is the exponential
distribution. Note also that the fast switching assumption between the two
states, $z_{\pm} \ll k_{\pm} = 1/\mu_{\pm}$, is necessary to use the expansion
in Eq. (\ref{e.rho(s).power-law(1,2)}) with $\sigma = z_{\pm}$. In addition to
this condition, we assume that $\mu_{\pm} z_{\pm} \gg
a_{\pm}z_{\pm}^{\alpha_{\pm}}$.
The auxiliary function $\tilde{\rho}^{\pm}(\sigma)$ [Eq.~(\ref{e.tilde.rho})] is then
given by
\begin{align}
  \label{e.tilde.rho(s).power-law(1,2)}
  \tilde{\rho}^{\pm}(\sigma) \underset{\sigma \to 0}{\simeq}
  \mu_{\pm} \sigma - a_{\pm} \sigma^{\alpha_{\pm}}  + o(\sigma^{\alpha_{\pm}}).
\end{align}

The function $\breve{f}^D (u; s | w^{\mathrm{eq}})$ can be given in a form
similar to Eq.~(\ref{e.f^R(u;s).general.2}). In fact, $\breve{f}^D (u; s |
w^{\mathrm{eq}})$ can be obtained just by replacing $\hat{\rho}^{\pm, 0}$ and
$p_{\pm}^0$ in Eq.~(\ref{e.f^R(u;s).general.2}) with $\hat{\rho}^{\pm,
  \mathrm{eq}}$ and $p_{\pm}^{\mathrm{eq}}$, respectively.
Furthermore, by using $\hat{\rho}^{\pm, \mathrm{eq}} (z_{\pm}) =
\tilde{\rho}^{\pm}(z_{\pm}) / (\mu_{\pm} z_{\pm})$ [see
Eq.~(\ref{e.init-deinsity-equilibrium})] and Eq.~(\ref{e.eq-ensemble-1}), we
rewrite Eq.~(\ref{e.f^R(u;s).general.2}) as
\begin{widetext}
  \begin{align}
    \label{e.f^R(u;s).power-law(1,2).2}
    \!\!\!\!\!
    \breve{f}^{D} (u; s | w^{\mathrm{eq}})
    =&
    \frac {\mu_+ + \mu_-}{\mu_+z_+ + \mu_-z_-}
    +
    \frac {\left(\frac {1}{z_+}- \frac {1}{z_-}\right)^2}{\mu_+ + \mu_-}
    \left[
    \frac {\mu_+\mu_-z_+z_-}{\mu_+z_+ + \mu_-z_-}
    -
    \frac
    {
      \tilde{\rho}^+(z_+)\tilde{\rho}^-(z_-)
    }
    {\tilde{\rho}^+(z_+) + \tilde{\rho}^-(z_-) - \tilde{\rho}^+(z_+) \tilde{\rho}^-(z_-)}
    \right].
  \end{align}
\end{widetext}
This expression for the equilibrium ensemble is valid for any sojourn-time PDFs
$\rho^{\pm}(\tau)$ with finite means $\mu_{\pm}$.

Now, substituting $ \tilde{\rho}^{\pm}(z_{\pm}) \simeq \mu_{\pm} z_{\pm} -
a_{\pm} z_{\pm}^{\alpha_{\pm}}$ into Eq.~(\ref{e.f^R(u;s).power-law(1,2).2}), we
have a formula for the power-law sojourn-time PDFs as
\begin{align}
  \label{e.f^R(u;s).power-law(1,2).3}
  \breve{f}^{D} &(u; s | w^{\mathrm{eq}})
  \simeq
  \frac {1}{s + \left\langle D \right\rangle_{\mathrm{eq}} u}
  \notag\\
  &+
  \frac
  {(\Delta Du)^2
    \left(
    a_- \mu_+^2
    z_-^{\alpha_- - 2}
    +
    a_+ \mu_-^2
    z_+^{\alpha_+ - 2}
    \right)
  }
  {\mu^3(s + \left\langle D \right\rangle_{\mathrm{eq}} u)^2},
\end{align}
where $\mu$ and $\Delta D$ are defined by $\mu := \mu_+ + \mu_-$, and $\Delta D
:= D_+ - D_-$, respectively.


\begin{figure*}[t]
  \begin{minipage}[bt]{8.5cm}
    \centerline{\includegraphics[width=8.0cm]{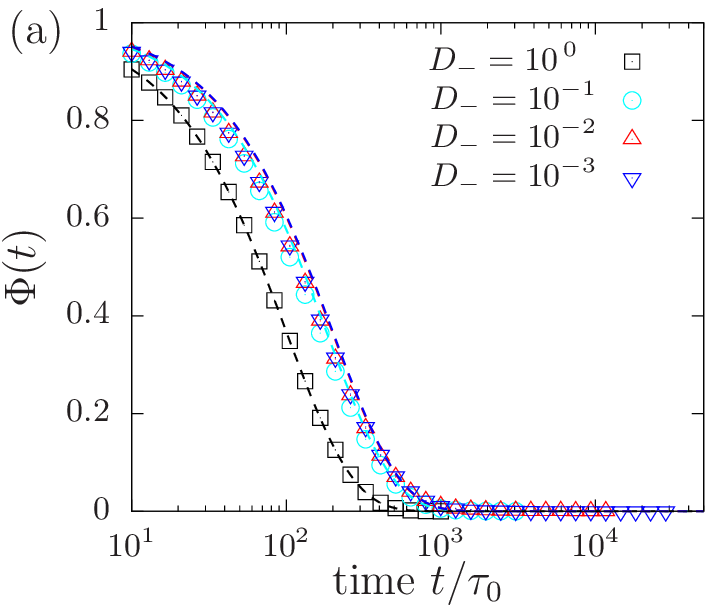}}
  \end{minipage}
  \begin{minipage}[bt]{8.5cm}
    \centerline{\includegraphics[width=8.0cm]{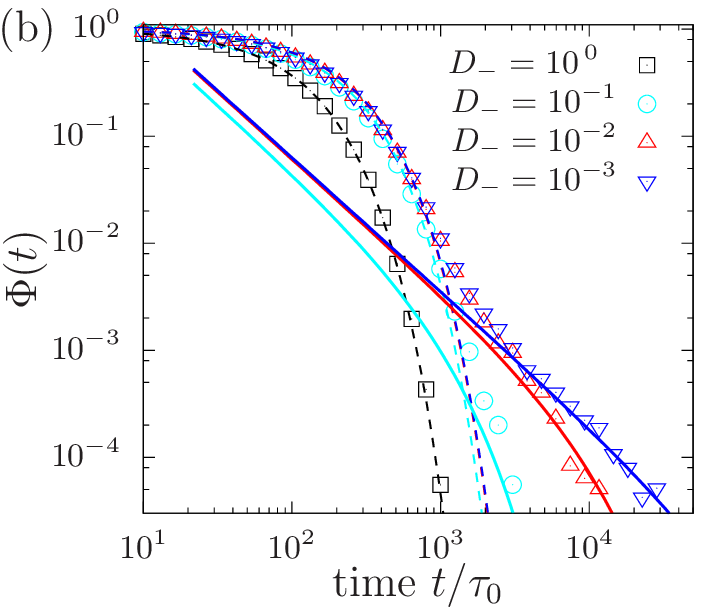}}
  \end{minipage}
  \caption{\label{f.power-law(1,2).neq} (Color online) Relaxation function
    $\Phi(t)$ as a function of time $t$ in (a) semi-log form and (b) log-log
    form. The initial ensemble for $D(t)$ is assumed to be the typical
    non-equilibrium ensemble $w^{\mathrm{neq}}$ with the initial fractions
    $p_{\pm}^0= p_{\pm}^{\mathrm{eq}}$. The other parameter values and the
    initial ensemble for $r(t)$ are exactly the same as those in
    Fig.~\ref{f.power-law(1,2).eq}. The dashed and solid lines are the
    theoretical predictions for the short and long time regimes, respectively
    [Eq.~(\ref{e.f^R(u;t).power-law(1,2).neq.short}) and
    Eq.~(\ref{e.f^R(u;t).power-law(1,2).neq.long})]. There are no fitting
    parameters.}
\end{figure*}


For $s \sim D_+ u$, the first term in
Eq.~(\ref{e.f^R(u;s).power-law(1,2).3}) is dominant over the second, and thus
the latter can be neglected.  Therefore, we have
\begin{equation}
  \label{e.f^R(u;s).power-law(1,2).eq.short}
  \breve{f}^{D} (u; s | w^{\mathrm{eq}})
  \simeq
  \frac {1}{s + \left\langle D \right\rangle_{\mathrm{eq}} u},
\end{equation}
or, by the Laplace inversion, we obtain
\begin{equation}
  \label{e.f^R(u;t).power-law(1,2).eq.short}
  \hat{f}^{D} (u; t | w^{\mathrm{eq}})
  \simeq
  e^{-\left\langle D \right\rangle_{\mathrm{eq}} ut}.
\end{equation}
Thus, at the short time scale $t \sim 1/D_+u$, the exponential relaxation is
observed.

However, for $s \ll D_+ u$, we have
$s+\left\langle D \right\rangle_{\mathrm{eq}} \sim \left\langle D
\right\rangle_{\mathrm{eq}}$ and $z_+ \sim D_+u$. The constant terms must be
exponentially small after the Laplace inversion, and thus, we have from
Eq.~(\ref{e.f^R(u;s).power-law(1,2).3})
\begin{equation}
  \label{e.f^R(u;s).power-law(1,2).eq.long}
  \breve{f}^{D} (u; s | w^{\mathrm{eq}})
  \simeq
  \frac {a_-}{\mu}
  \left(\frac {\mu_+ \Delta D}{\mu \left\langle D \right\rangle_{\mathrm{eq}}}\right)^2
  z_-^{\alpha_- - 2}.
\end{equation}
By the Laplace inversion, we have the following expression in the time domain $t
\gg 1/D_+u$
\begin{equation}
  \label{e.f^R(u;t).power-law(1,2).eq.long}
  \hat{f}^{D} (u; t | w^{\mathrm{eq}})
  \simeq
  \frac {a_-}{\mu}
  \left(\frac {\mu_+ \Delta D}{\mu \left\langle D \right\rangle_{\mathrm{eq}}}\right)^2
  \frac {e^{-D_- ut} t^{1-\alpha_-}}{|\Gamma(2-\alpha_-)|}.
\end{equation}
Thus, in the case $1< \alpha_-<2$, the relaxation function exhibits the
exponential decay at the short time scale
[Eq.~(\ref{e.f^R(u;t).power-law(1,2).eq.short})], whereas it exhibits the
power-law decay with an exponential cutoff at the long time scale
[Eq.~(\ref{e.f^R(u;t).power-law(1,2).eq.long})].
As shown in Fig.~\ref{f.power-law(1,2).eq}, these theoretical results are
consistent with numerical simulations. It is possible to show that
Eqs.~(\ref{e.f^R(u;s).power-law(1,2).eq.short})--
(\ref{e.f^R(u;t).power-law(1,2).eq.long}) are valid also for $\alpha_{\pm} > 2$
(it should be assumed that $\alpha_{-} < \alpha_{+}$ and that $\alpha_-$ is not
an integer). Thus, the power-law dacay $1/t^{1-\alpha_-}$ with any values of
$\alpha_- > 1$ is possible in principle. However, for $\alpha_- > 2$, the decay
is relatively fast, and therefore it might be difficult to observe the predicted
power-law decay behavior even in numerical simulations.

\subsubsection {Non-equilibrium ensemble}

The relaxation function for the typical non-equilibrium ensemble
$w^{\mathrm{neq}}(\tau_1) = p_{+}^{0} \rho^{+}(\tau_1) + p_{-}^{0}
\rho^{-}(\tau_1)$ with $p_+^0 + p_-^0 = 1$ can be derived in the same way as the
case of the equilibrium ensemble. By replacing $\hat{\rho}^{\pm, 0}(z_{\pm})$ in
Eq.~(\ref{e.f^R(u;s).general.2}) with $\hat{\rho}^{\pm}(z_{\pm})$,
and substituting Eqs.~(\ref{e.rho(s).power-law(1,2)}) and
(\ref{e.tilde.rho(s).power-law(1,2)}) into the resulting equation,
we obtain
\begin{widetext}
  \begin{align}
    \label{e.f^R(u;s).power-law(1,2).noneq.2}
    \breve{f}^{D} & (u; s | w^{\mathrm{neq}})
    =
    \frac {\mu_+ + \mu_-}{\mu_+z_+ + \mu_-z_-}
    \notag\\[0.1cm]
    +&
    \frac {z_+ - z_-}{z_+ z_-}
    \left[
    \frac {p_-^0\mu_+z_+ - p_+^0 \mu_-z_-}{\mu_+z_+ + \mu_-z_-}
    +
    \frac
    {
      p_+^0\tilde{\rho}^-(z_-) - p_-^0\tilde{\rho}^+(z_+)
      +
      (p_-^0 - p_+^0) \tilde{\rho}^+(z_+)\tilde{\rho}^-(z_-)
    }
    {\tilde{\rho}^+(z_+) + \tilde{\rho}^-(z_-) - \tilde{\rho}^+(z_+) \tilde{\rho}^-(z_-)}
    \right].
  \end{align}
\end{widetext}
Note that this expression for the typical non-equilibrium ensemble
$w^{\mathrm{neq}}$ is valid for any sojourn-time PDF $\rho^{\pm}(\tau)$
with finite means $\mu_{\pm}$.


Now, substituting $ \tilde{\rho}^{\pm}(z_{\pm}) \simeq \mu_{\pm} z_{\pm} -
a_{\pm} z_{\pm}^{\alpha_{\pm}}$, we have a formula for the power-law
sojourn-time PDFs as
\begin{align}
  \breve{f}^{D} (u; s | w^{\mathrm{neq}})&
  \simeq
  \frac {1}{s + \left\langle D \right\rangle_{\mathrm{eq}} u}
  \notag\\[0.1cm]
  \label{e.f^R(u;s).power-law(1,2).noneq.3}
  +&
  \frac
  {\Delta D u
    \left(
    \mu_- a_+ z_+^{\alpha_+-1}
    -
    \mu_+ a_- z_-^{\alpha_--1}
    \right)
  }
  {\mu^2\left(s + \left\langle D \right\rangle_{\mathrm{eq}} u\right)^2}.
\end{align}
%

\begin{figure*}[t]
  \begin{minipage}[bt]{8.5cm}
    \centerline{\includegraphics[width=8.0cm]{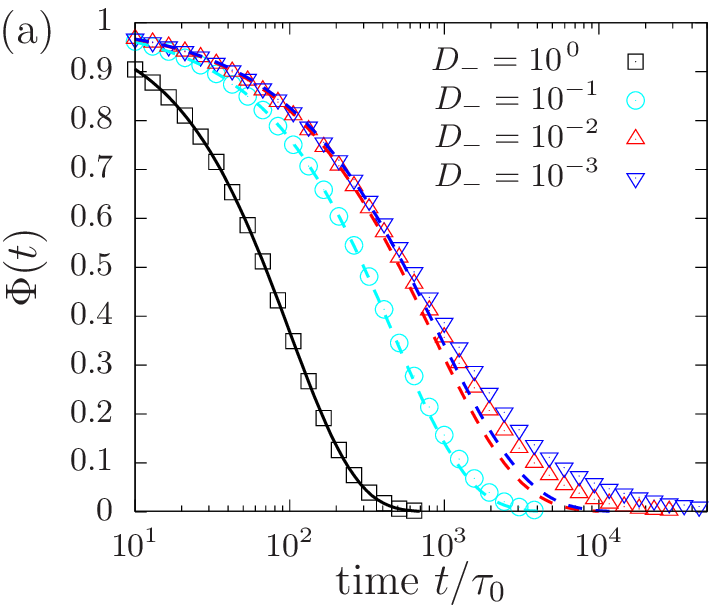}}
  \end{minipage}
  \begin{minipage}[bt]{8.5cm}
    \centerline{\includegraphics[width=8.0cm]{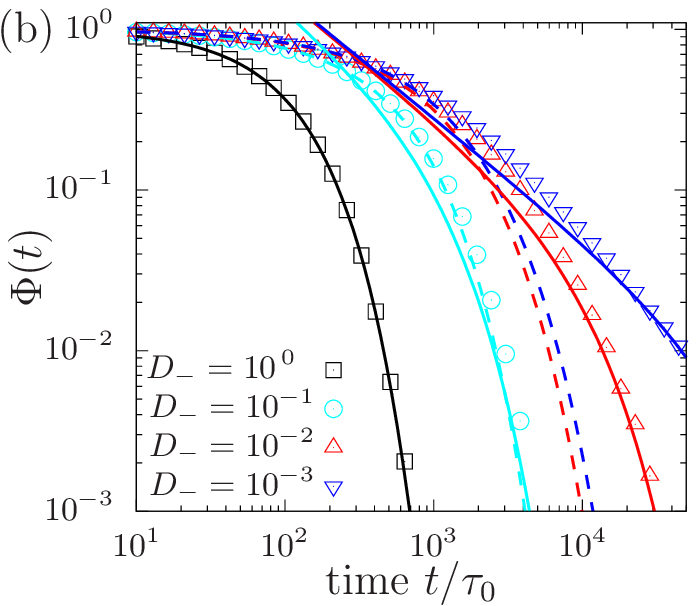}}
  \end{minipage}
  \caption{\label{f.power-law(0,1).neq} (Color online) Relaxation function
    $\Phi(t)$ as a function of time $t$ in (a) semi-log form and (b) log-log
    form. The initial ensemble for $D(t)$ is assumed to be the non-equilibrium
    ensemble $w^{\mathrm{neq}}$ with the initial fractions $p_{\pm}^0= 0.5$.
    The scaling exponent $\alpha_{-}$ of the power law distribution
    $\rho^{-}(\tau)$ is set as $\alpha_-= 0.75$. In contrast, $\rho^{+}(\tau)$
    is assumed to be the exponential distribution with $\mu_+ = 2 \tau_0$.  The
    other parameter values and the initial ensemble for $r(t)$ are exactly the
    same as those in Fig.~\ref{f.power-law(1,2).eq}. The dashed and solid lines
    are the theoretical predictions for the short and long time regimes,
    respectively [Eqs.~(\ref{e.f^R(u;t).power-law(0,1).short}) and
    (\ref{e.f^R(u;t).power-law(0,1).long})]. There is no fitting parameters.}
\end{figure*}

%
For $s \sim D_+ u$, the first term in
Eq.~(\ref{e.f^R(u;s).power-law(1,2).noneq.3}) is dominant over the second, and
thus the latter can be neglected. Therefore, we have
\begin{equation}
  \label{e.f^R(u;s).power-law(1,2).neq.short}
  \breve{f}^{D} (u; s | w^{\mathrm{neq}})
  \simeq
  \frac {1}{s + \left\langle D \right\rangle_{\mathrm{eq}} u}
\end{equation}
or, by the Laplace inversion, we obtain
\begin{equation}
  \label{e.f^R(u;t).power-law(1,2).neq.short}
  \hat{f}^{D} (u; t | w^{\mathrm{neq}})
  \simeq
  e^{-\left\langle D \right\rangle_{\mathrm{eq}} ut}.
\end{equation}
Thus, at the short time scale $t \sim 1/D_+u$, the exponential relaxation, which
is exactly the same as that in the equilibrium ensemble
[Eq.~(\ref{e.f^R(u;t).power-law(1,2).eq.short})], is observed.

However, for $s \ll D_+ u$, the constant terms can be neglected because they are
exponentially small after the Laplace inversion, and thus we have
\begin{equation}
  \label{e.f^R(u;s).power-law(1,2).neq.long}
  \breve{f}^{D} (u; s | w^{\mathrm{neq}})
  \simeq
  -
  \frac
  {a_- \mu_+\Delta D }
  {\mu^2 \left\langle D \right\rangle_{\mathrm{eq}}^2 u}
  z_-^{\alpha_- - 1},
\end{equation}
or by the Laplace inversion, we obtain a power-law decay again for the long time
regime $t \gg 1/D_+u$ as
\begin{equation}
  \label{e.f^R(u;t).power-law(1,2).neq.long}
  \hat{f}^{D} (u; t | w^{\mathrm{neq}})
  \simeq
  \frac {a_-\mu_+ \Delta D}{\mu^2 \left\langle D \right\rangle^2_{\mathrm{eq}} u}
  \frac {e^{-D_- ut} t^{-\alpha_-}}{|\Gamma(1-\alpha_-)|}.
\end{equation}
As shown in Fig.~\ref{f.power-law(1,2).neq}, these theoretical results are
consistent with numerical simulation. These results
[Eqs.~(\ref{e.f^R(u;s).power-law(1,2).neq.short})--~(\ref{e.f^R(u;t).power-law(1,2).neq.long})]
are valid also for $\alpha_{\pm} > 2$ (it should be assumed that $\alpha_{-} <
\alpha_{+}$ and that $\alpha_-$ is not an integer).

As in the case of the equilibrium ensemble, the relaxation for the
non-equilibrium ensemble is given by the exponential decay at the short time
scale [Eq.~(\ref{e.f^R(u;t).power-law(1,2).neq.short})], whereas power-law decay
with an exponential cutoff at the long time scale
[Eq.~(\ref{e.f^R(u;t).power-law(1,2).neq.long})]. Thus, the behavior of both
ensembles are qualitatively similar except that the power-law decay is slower
for the equilibrium ensemble than for the non-equilibrium ensemble. In fact, the
power-law index for equilibrium ensemble is greater by one than that for the
non-equilibrium ensemble [See Eqs.~(\ref{e.f^R(u;t).power-law(1,2).eq.long}) and
(\ref{e.f^R(u;t).power-law(1,2).neq.long})]. This difference is due to the fact
that the power-law index of the initial sojourn-time PDF for the equilibrium
ensemble is smaller by one than that for the non-equilibrium ensemble [See
Eq.~(\ref{e.equilibrium.power-law})]. This difference of power-law exponents in
the initial sojourn-time PDFs often provides a remarkable initial-ensemble
dependence of physical observables such as the correlation function and the
diffusion coefficient \cite{akimoto07, akimoto18}.

\subsection {Power-law distribution: $0<\alpha_-<1$}\label{s.case-studies.3}

Finally, let us study the case in which $0<\alpha_-<1$. In this case, the
equilibrium distribution $\rho^{-, \mathrm{eq}}(\tau)$ no longer exists, because
the mean value of $\rho^{-}(\tau)$ diverges. Hence, we consider only the
non-equilibrium ensemble $w^{\mathrm{neq}}$. On the other hand, $\alpha_+$ is
assumed to be $1< \alpha_+ \leq 2$ (For $0 < \alpha_+ < 1$, we can have similar
results with different scaling exponents. But the behavior for these two cases
are qualitatively the same, and thus here we focus only on the case $1< \alpha_+
\leq 2$). In addition to the assumptions for $\hat{\rho}^{+}(z_{+})$ used in the
previous subsection, we further assume that $a_- z_-^{\alpha_-} \ll 1$; this
condition is necessary for the expansion given in
Eq.~(\ref{e.rho(s).asymptotic.alpha<1}).

The double Laplace transform $\breve{f}^D (u; s | w^{\mathrm{neq}})$ is obtained
by replacing $\hat{\rho}^{\pm, 0}(z_{\pm})$ in Eq.~(\ref{e.f^R(u;s).general.2})
with $\hat{\rho}^{\pm}(z_{\pm})$. Substituting
$\hat{\rho}^{\pm} (z_{\pm}) \simeq 1$,
$\tilde{\rho}^{+}(z_+) \simeq \mu_{+} z_+$ and
$\tilde{\rho}^{-}(z_-) \simeq a_{-} z_-^{\alpha_{-}}$ into this equation, we
have for $s \sim D_+u$
\begin{equation}
  \label{e.f^R(u;s).power-law(0,1).short}
  \breve{f}^{D} (u; s | w^{\mathrm{neq}})
  \simeq
  \frac {1}{z_-}
  -
  \frac {\mu_+}{a_-} \frac {\Delta D u}{z_-^{1+\alpha_-}}.
\end{equation}
Strictly speaking, higher order terms should have been taken into account in
this calculation, but the final result is unchanged because the higher order
terms are canceled out.  The Laplace inversion of
Eq.~(\ref{e.f^R(u;s).power-law(0,1).short}) gives
\begin{align}
  \hat{f}^{D} (u; t | w^{\mathrm{neq}})
  &\simeq
  e^{- D_-ut}
  \left[
  1
  -
  \frac {\mu_+}{a_-} \frac {\Delta Du } {\Gamma(1+\alpha_-)} t^{\alpha_-}
  \right]
  \notag\\[0.1cm]
  \label{e.f^R(u;t).power-law(0,1).short}
  &\simeq
  e^{- D_-ut - \frac {\mu_+}{a_-} \frac {\Delta Du} {\Gamma(1+\alpha_-)} t^{\alpha_-}}.
\end{align}
Thus, the relaxation function is given in a stretched-exponential form.

However, for $s \ll D_+ u$, we have the following formula from the equation
obtained by replacing $\hat{\rho}^{\pm, 0}(z_{\pm})$ in
Eq.~(\ref{e.f^R(u;s).general.2}) with $\hat{\rho}^{\pm}(z_{\pm})$
\begin{equation}
  \label{e.f^R(u;s).power-law(0,1).long}
  \breve{f}^{D} (u; s | w^{\mathrm{neq}})
  \simeq
  \frac {\Delta D}{D_+}
  \frac {a_-}{\mu_+}
  \frac {z_-^{\alpha_--1}}{D_+ u},
\end{equation}
where we also assumed that $z_-^{\alpha_-} \ll D_{+}u$ (if this condition is not
satisfied, the power-law regime vanishes).  The Laplace inversion gives
\begin{equation}
  \label{e.f^R(u;t).power-law(0,1).long}
  \hat{f}^{D} (u; t | w^{\mathrm{neq}})
  \simeq
  \frac {\Delta D}{D_+^2 u}
  \frac {a_-}{\mu_+}
  \frac {e^{-D_-  ut}}{\Gamma(1-\alpha_-) t^{\alpha_-}}.
\end{equation}
Thus, in the case $0< \alpha_-<1$, the relaxation function exhibits the
stretched-exponential decay at the short time scale, whereas the power-law decay
with an exponential cutoff at the long time scale. As shown in
Fig.~\ref{f.power-law(0,1).neq}, these theoretical results are consistent with
numerical simulation.

\section {Discussion}\label{s.discussion}

Recently, Brownian motion with fluctuating diffusivity has been investigated
intensively. To characterize such processes, various kinds of correlation
functions are used. One of the most frequently used functions might be an
ergodicity breaking parameter, which is a fourth order correlation function in
terms of the position vector \cite{uneyama12, uneyama15, uneyama19, miyaguchi16,
  miyaguchi17, cherstvy13, cherstvy16, metzler14}. For the case of the LEFD, it
is shown that the fluctuating diffusivity cannot be detected with the second
order correlation including the MSD \cite{miyaguchi17}.


In this work, we have studied the relaxation function $\hat{f}^R(u; t | w^0)$.
This quantity can be interpreted as a second order position correlation function
[Eq.~(\ref{e.def.relaxation-func})] for the OUFD, and as the self-intermediate
scattering function [Eq.~(\ref{e.Fs(k,t)})] for the LEFD. This correlation
function characterizes the fluctuating diffusivity in the sense that it is a
Laplace transformation of the integral diffusivity $\int_{0}^{t}dt' D(t')$.


Under the assumption that the system has two diffusive states, we have derived a
general formula [Eq.~(\ref{e.f^R(u;s).general.2})] for the relaxation function
$\hat{f}^R(u; t | w^0)$ in terms of the sojourn-time PDFs $\rho^{\pm}(\tau)$. In
particular, if the sojourn-time PDF of the slow state is given by a power-law
distribution with the index $\alpha_-$, the relaxation function at the short
time scales shows the exponential relaxation for $1< \alpha_- <2$, or the
stretched-exponential relaxation for $0< \alpha_- <1$. In contrast, at the long
time scales, the relaxation function shows a power law decay with an exponential
cutoff.


The fact that our model can reproduce the stretched-exponential relaxation
function is quite interesting. The stretched-exponential type relaxation
behavior has been widely observed in various systems, and generally interpreted
as the signature of a heterogeneity.  Various models have been proposed to
interpret the microscopic origin of the stretched-exponential relaxation
\cite{palmer84, odagaki90, ngai93, phillips96, metzler00}. In most of such
models, power-law type distributions for some physical quantities such as the
waiting time are assumed.

Our model also assumes the power-law distribution for the sojourn time, and
thus, in this aspect, it is similar to the conventional models. In particular,
it would be worth mentioning a relation of the present model with the
quenched-trap model studied in Ref.~\cite{odagaki90}. Statistical properties of
the quenched-trap model with the spatial dimension $d > 2$ is asymptotically the
same as those of the continuous-time random walks \cite{machta85}. On the other
hand, the LEFD also reduces to the continuous-time random walk in the limit
$\mu_+ \to 0$ with $D_+ \mu_+ = \mathrm{const}$ and $D_- = 0$
\cite{uneyama15}. Thus, the generation mechanism of the stretched-exponential
relaxation (as well as the anomalous diffusion \cite{miyaguchi16}) in the
present model may be similar to that in the quenched-trap model
\cite{odagaki90}.


In spite of these similarities between our model and previously studied systems,
however, our model might have some advantages in that its dynamics is described
by simple equations [Eqs.~\eqref{e.ou_proc.w.fluctuating-diffusivity} and
\eqref{e.lengeven-eq.w.fluctuating-diffusivity}]. This is in contrast to some
conventional models, where the dynamic equation and the time evolution rule are
not expressed in clear and simple equations. We expect that the combination of
our model and experimental relaxation data can be employed to interpret
molecular-level dynamics of heterogeneous systems. In fact, a model similar to
ours was utilized to reproduce results of numerical simulations for glassy
systems and supercooled liquids \cite{chaudhuri07, helfferich18}.

Moreover, the stretched-exponential relaxation is obtained in our model when the
equilibrium distribution $\rho^{-,\mathrm{eq}}(\tau)$ does not exist. This means
that, the system is not in equilibrium, if it shows the stretched exponential
type relaxation. This seems to be consistent with a naive expectation that the
dynamic heterogeneity is a precursor of a glass transition which is clearly in
non-equilibrium.

As we mentioned, the stretched-exponential relaxation is observed ubiquitously
in complex and highly concentrated systems , but its origin is still unclear
\cite{phillips96}. The stretched-exponential relaxation is often expressed as
$e^{- (t / \bar{\tau})^{\beta}}$ where $\bar{\tau}$ and $\beta$ are constants;
$\beta$ is typically around $0.5$ (experimental values are about $0.4 \lesssim
\beta \lesssim 0.6$).  In our model, this exponent is simply given as $\beta =
\alpha_{-}$, and therefore it reflects the tail of the sojourn-time PDF of the
slow state at the long time region. This result implies that the power-law
sojourn-time PDF for the slow state plays a fundamental role in emergence of the
stretched-exponential relaxation. The power-law sojourn-time PDF, which is
assumed in our model, is also generally observed in crowded systems
\cite{doliwa03, helfferich14a}. Thus, our model might bridge a gap between these
commonly observed phenomena: the power-law sojourn times and
stretched-exponential relaxation.

Although it seems that the fast state plays a minor role, its property affects
the characteristic relaxation time; in our model, the characteristic relaxation
time $\bar{\tau}$ is given as $\bar{\tau} = [\mu_{+} \Delta D u / a_{-} \Gamma(1
+ \alpha_{+})]^{1/\alpha_{-}}$. Thus, the characteristic time scale depends on
various parameters such as the sojourn-time PDF of the fast state and the two
diffusion coefficients.

\section* {Acknowledgments}\label{acknowledgment}

T.M. was supported by Grant-in-Aid (KAKENHI) for Scientific Research C (Grand
No. JP18K03417). T.U. was supported by Grant-in-Aid (KAKENHI) for Scientific
Research C (Grant No. JP16K05513).  T.A. was supported by Grant-in-Aid (KAKENHI)
for Scientific Research B (Grand No. JP16KT0021), and Scientific Research C
(Grant No. JP18K03468).

\appendix {}
\section {Equilibrium ensemble}\label{s.equilibrium-ensemble}

As explained in Sec.~\ref{s.model}, the initial ensemble is fully specified by
the PDF of the first sojourn time $\tau_1$ [Recall that the initial distribution
of the position $\bm{r}(0)$ is unimportant in investigating the relaxation
function $\Phi(t)$].
We derive an explicit expression for the Laplace transform of the equilibrium
ensemble
$\hat{\rho}^{\pm, \mathrm{eq}}(\sigma):=\int_{0}^{\infty} d\tau e^{-\tau\sigma}
\rho^{\pm, \mathrm{eq}}(\tau)$.


The equilibrium ensemble can be derived from a forward-recurrence-time PDF. A
forward-recurrence time is a sojourn time until the next transition from some
elapsed time $t' > 0$ \cite{godrche01}. Here, we study the
forward-recurrence-time PDF $w_\pm(\tau;t'| w^0)$, from which we obtain the
equilibrium ensemble by taking a limit $t' \to \infty$.


\begin{figure}[t!]
  \centerline{\includegraphics[width=8.5cm]{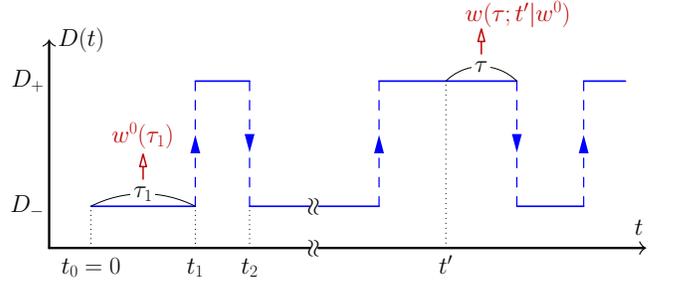}}
  \caption{\label{f.forward-recurrence} (Color online)Schematic illustration of
    the forward-recurrence time $\tau$.  A realization started from the initial
    ensemble $w^0(\tau_1)$ is displayed as an example. The first sojourn time
    $\tau_1$ follows $w^0(\tau_1) = w_+^0(\tau_1) + w_-^0(\tau_1)$ [See
    Eq.~(\ref{e.def.genelal.emsenble})]. Then, the sojourn time $\tau$ at time
    $t'$ until the next transition follows
    $w(\tau;t'| w^0):= w_+(\tau;t'| w^0) + w_-(\tau;t'| w^0)$.}
\end{figure}

\subsection {PDF for forward-recurrence time}

The forward-recurrence-time PDF $w_\pm(\tau;t'| w^0) d\tau$ is a joint
probability that %
(i) the state is $\pm$ at time $t'$, and %
(ii) the sojourn time from $t'$ until the next transition is in an interval
$[\tau, \tau+d\tau]$, given that the process starts with $w^0(\tau_1)$ at $t=0$.
Here, $\tau$ is called the forward-recurrence time, and it is illustrated in
Fig.~\ref{f.forward-recurrence}.

To obtain the forward-recurrence-time PDF, we further define the following PDFs
for $n=0,1,\dots$:
\begin{equation}
  w_{\pm,n}(\tau; t' | w^0) :=
  \left\langle
  \delta \left(\tau - (t_{n+1} - t')\right)
  I \left( t_n < t' < t_{n+1} \right)
  \delta_{\pm}
  \right\rangle,
\end{equation}
where $w_{\pm, n}(\tau; t'| w^0)d\tau$ is a joint probability that %
(i) the state is $\pm$ at $t=0$, %
(ii) there are $n$ transitions until time $t'$, and %
(iii) the sojourn time at time $t'$ is in an interval $[\tau, \tau+d\tau]$, %
given that the process starts with $w^0(\tau_1)$ at $t=0$.

The double Laplace transforms
$\breve{w}_{\pm,n} (\sigma; s| w^0) := \int_0^{\infty} dt' e^{-t's}
\int_0^{\infty}d\tau e^{-\tau\sigma}w_{\pm,n}(\tau; t' | w^0)$ with respect to
$\tau$ and then $t'$ result in ($\tau \leftrightarrow \sigma$ and
$t' \leftrightarrow s$)
\begin{align}
  \breve{w}_{\pm,n} (\sigma; s| w^0) &=
  \left\langle
  \frac {\delta_{\pm} e^{-t_ns}}{s-\sigma}
  \left(
  e^{-\sigma\tau_{n+1}}-e^{-s\tau_{n+1}}
  \right)
  \right\rangle
  \notag\\[0.0cm]
  \label{e.w_pm,n(sigma,s)}
  &\hspace*{-2.1cm}=
  \begin{cases}
    \displaystyle
    \frac {
      \left\langle
      \delta_{\pm} e^{-\sigma\tau_1}
      \right\rangle
      -
      \left\langle
      \delta_{\pm} e^{-s\tau_1}
      \right\rangle
    }
    {s-\sigma}
    & (n=0),\\[0.20cm]
    \displaystyle
    \left\langle \delta_{\pm} e^{-st_n} \right\rangle
    \frac
    {\left\langle
      e^{-\sigma\tau_{n+1}}
      \right\rangle_{\pm}
      -
      \left\langle
      e^{-s\tau_{n+1}}
      \right\rangle_{\pm}}
    {s-\sigma}
    & (n=1,2,\dots),
  \end{cases}
\end{align}
where $\left\langle \dots \right\rangle_{\pm}$ is a conditional average for a
given initial state $\pm$.  The ensemble averages in the above equation can be
expressed with the Laplace transforms of the initial ensemble $\hat{w}_{\pm}^0
(s)$ and the sojourn-time PDFs $\hat{\rho}^{\pm}(s)$.  In fact, we have
\begin{align}
  \left\langle \delta_{\pm} e^{-st_{2n}} \right\rangle
  &=
  \left\langle \delta_{\pm}  \right\rangle
  \left\langle e^{-s\tau_1} \right\rangle_{\pm}
  \left\langle e^{-s\tau_2} \right\rangle_{\pm}
  \times\dots\times
  \left\langle e^{-s\tau_{2n}} \right\rangle_{\pm}
  \notag\\[0.1cm]
  \label{e.app.<delta.exp>_2n}
  &=
  \hat{w}_{\pm}^0 (s) \hat{\rho}^{n-1}(s) \hat{\rho}^{\mp} (s),
  \\[0.3cm]
  \left\langle \delta_{\pm} e^{-s\tau_{2n-1}} \right\rangle
  &=
  \left\langle \delta_{\pm} \right\rangle
  \left\langle e^{-s\tau_1} \right\rangle_{\pm}
  \left\langle e^{-s\tau_2} \right\rangle_{\pm}
  \times\dots\times
  \left\langle  e^{-s\tau_{2n-1}} \right\rangle_{\pm}
  \notag\\[0.1cm]
  \label{e.app.<delta.exp>_2n-1}
  &=
  \hat{w}_{\pm}^0 (s) \hat{\rho}^{n-1}(s),
\end{align}
for $n=1, 2,\cdots$, where $\hat{\rho}(s) := \hat{\rho}^+(s)\hat{\rho}^-(s)$ and
$\left\langle \dots \right\rangle_{\pm}$ is a conditional average under the
condition that the initial state is $\pm$.  Then, using
Eqs.~(\ref{e.app.<delta.exp>_2n}) and (\ref{e.app.<delta.exp>_2n-1}) in
Eq.~(\ref{e.w_pm,n(sigma,s)}), we have
\begin{align}
  \breve{w}_{\pm,0} (\sigma; s | w^0) &=
  \dfrac {\hat{w}_{\pm}^0 (\sigma) - \hat{w}_{\pm}^{0} (s)}{s-\sigma},
  \\[0.0cm]
  \breve{w}_{\pm,2n} (\sigma; s | w^0) &=
  \hat{w}^0_{\pm}(s)\hat{\rho}^{n-1} (s) \hat{\rho}^{\mp}(s)
  \dfrac {\hat{\rho}^{\pm} (\sigma) - \hat{\rho}^{\pm} (s)} {s-\sigma},
  \\[0.0cm]
  \breve{w}_{\pm,2n-1} (\sigma; s | w^0) &=
  \hat{w}^0_{\pm}(s)\hat{\rho}^{n-1} (s)
  \dfrac {\hat{\rho}^{\mp} (\sigma) - \hat{\rho}^{\mp} (s)} {s-\sigma},
\end{align}
for $n=1,2,\dots$.

Because the Laplace transform of the forward-recurrence-time PDF, $w_\pm(\tau;t'
| w^0)$, is given by $\breve{w}_{\pm} (\sigma; s | w^0) =
\sum_{n=0}^{\infty}[\breve{w}_{\pm,2n} (\sigma; s | w^0 ) + \breve{w}_{\mp,2n+1}
(\sigma; s | w^0)]$, we have
\begin{align}
  \label{e.app.w(u;s).general}
  \breve{w}_{\pm}(\sigma; s | w^0) &=
  \frac {\hat{w}^0_{\pm} (\sigma) - \hat{w}^0_{\pm} (s)}{s-\sigma}\notag\\[0.0cm]
  +&
  \frac {\hat{w}^0_{\pm}(s) \hat{\rho}^{\mp}(s) + \hat{w}^0_{\mp}(s)}{1 - \hat{\rho} (s)}
  \frac {\hat{\rho}^{\pm} (\sigma) - \hat{\rho}^{\pm} (s)} {s-\sigma}.
\end{align}
Equation (\ref{e.app.w(u;s).general}) is a general expression of the
forward-recurrence-time PDF for the alternating renewal processes.

\subsection {Equilibrium ensemble}

If both $\mu_+$ and $\mu_-$ exist, it follows from
Eq.~(\ref{e.app.w(u;s).general}) that
\begin{equation}
  \label{e.app.eq-init-ensenble}
  \lim_{t' \to \infty} \hat{w}_{\pm}(\sigma; t'  | w^0)
  =
  \lim_{s \to 0} s \breve{w}_{\pm}(\sigma; s  | w^0)
  =
  p_{\pm}^{\mathrm{eq}}\frac {1 - \hat{\rho}^{\pm} (\sigma)}{\mu_{\pm} \sigma}.
\end{equation}
Since the above limit $t' \to \infty$ is equivalent to putting the start time of
the process to $-\infty$, the limitting function should be equivalent to the
equilibrium ensemble $\hat{w}^{\mathrm{eq}}_{\pm} (\sigma)$:
\begin{equation}
  \label{e.eq-ensemble-2}
  \hat{w}_{\pm}^{\mathrm{eq}}(\sigma) :=
  p_{\pm}^{\mathrm{eq}}\hat{\rho}^{\pm,\mathrm{eq}}(\sigma)
\end{equation}
where $\hat{\rho}_{}^{\pm,\mathrm{eq}}(\sigma)$ is defined as
\begin{equation}
  \label{e.init-deinsity-equilibrium}
  \hat{\rho}^{\pm, \mathrm{eq}}(\sigma)
  =
  \frac {1 - \hat{\rho}^{\pm}(\sigma)}{\mu_{\pm}\sigma},
\end{equation}
Hence, the equilibrium PDF exists only if $\mu_{\pm}$ is finite as explained in the
main text.  Note also that the equilibrium ensemble
[Eq.~(\ref{e.app.eq-init-ensenble})] does not depend on the initial ensemble
$w^0$, as expected.

\subsubsection {Exponential distribution}
If the sojourn-time PDFs are given by the exponential distributions
\begin{equation}
  \label{e.exp-dist}
  \rho^{\pm} (\tau)
  =
  \frac {1}{\mu_{\pm}} \exp\left(- \frac {\tau}{\mu_{\pm}}\right)
  =
  k_{\pm} \exp\left(- k_{\pm}{\tau}\right),
\end{equation}
we have its Laplace transforms ($\tau \leftrightarrow \sigma$) as
\begin{equation}
  \label{e.laplace.exp-dist}
  \hat{\rho}^{\pm} (\sigma)
  = \frac {1}{1 + \mu_{\pm} \sigma}
  = \frac {k_{\pm}}{k_{\pm} + \sigma}.
\end{equation}
In this case, the equilibrium PDF [Eq.~(\ref{e.init-deinsity-equilibrium})] is
equivalent to $\rho^{\pm} (\tau)$, i.e., it is easy to check
\begin{equation}
  \label{e.laplace.exp-dist.stationality}
  \hat{\rho}^{\pm, \mathrm{eq}} (\sigma)
  =
  \frac {1 - \hat{\rho}^{\pm}(\sigma)}{\mu_{\pm}\sigma}
  =
  \hat{\rho}^{\pm} (\sigma).
\end{equation}
Thus, the equilibrium ensemble is given by $p_+^{\mathrm{eq}} \rho_+(\tau_1) +
p_-^{\mathrm{eq}} \rho_-(\tau_1)$.

More generally, the forward-recurrence-time PDF for the exponential sojourn
times is calculated as follows. Here, let us assume that the initial ensemble is
the typical non-equilibrium ensemble, i.e.,
$w_{\pm}^{\mathrm{neq}}(\tau_1) = p_+^0 \rho_+(\tau_1) + p_-^0 \rho_-(\tau_1)$,
where $p_{\pm}^0$ are the initial fractions with the constraint
$p_+^0 + p_-^0 =1$. Then, after some straightforward calculations, the
forward-recurrence-time PDF [Eq.~(\ref{e.app.w(u;s).general})] is given as
\begin{equation}
  \label{e.forward-reccurence-PDF.exponential.w(sigma;s)}
  \breve{w}_{\pm}(\sigma; s| w^{\mathrm{neq}})
  =
  \frac {1}{1+\mu_{\pm} \sigma}
  \left[
  \frac {p_{\pm}^{\mathrm{eq}}}{s}
  +
  \frac {p_{\pm}^0 - p_{\pm}^{\mathrm{eq}}}{s+ \frac {\mu_+ + \mu_-}{\mu_+ \mu_-}}
  \right].
\end{equation}
The double Laplace inversions give
\begin{equation}
  \label{e.forward-reccurence-PDF.exponential.w(tau;t')}
  w_{\pm}(\tau; t'| w^{\mathrm{neq}})
  =
  p_{\pm}(t')
  \frac {e^{-\tau/\mu_{\pm}}}{\mu_{\pm}},
\end{equation}
where $p_{\pm}(t')$ are the fractions at time $t'$, and are given as
\begin{equation}
  \label{e.forward-reccurence-PDF.exponential.p(t')}
  p_{\pm} (t')
  :=
  p_{\pm}^{\mathrm{eq}}
  +
  \left(p_{\pm}^0 - p_{\pm}^{\mathrm{eq}}\right)
  e^{-\frac {\mu_+ + \mu_-}{\mu_+ \mu_-}t}.
\end{equation}
From Eqs.~(\ref{e.forward-reccurence-PDF.exponential.w(tau;t')}) and
(\ref{e.forward-reccurence-PDF.exponential.p(t')}), the forward-recurrence-time
PDF at $t'$ is also represented as
\begin{equation}
  \label{e.forward-reccurence-PDF.exponential.rho(tau)p(t')}
  p_{+}(t') \rho^+(\tau) + p_{-}(t') \rho^-(\tau),
\end{equation}
Thus, the forward-recurrence-time PDF with respect to the exponential
distributions are also given by the exponential distributions, and only the
fractions change with time. This is an outcome of the fact that the exponential
distribution is Markovian.  In addition, the fractions converge to the
equilibrium fractions $p_{\pm}^{\mathrm{eq}}$ exponentially.

\subsubsection {Power law distribution}

The power law distribution is given by
\begin{equation}
  \label{e.rho(t)}
  \rho^\pm (\tau)
  \underset{\tau\to\infty}{\simeq}
  \frac {a_{\pm}}{|\Gamma (-\alpha_{\pm})|\tau^{1+\alpha_\pm}},
\end{equation}
where $a_{\pm}$ is a scale factor and $\Gamma (-\alpha_{\pm})$ is the gamma
function. Asymptotic forms of the Laplace transforms $\hat{\rho}_{\pm} (\sigma)
:= \int_0^{\infty} d\tau\, e^{-\tau \sigma} \rho_{\pm}(\tau)$ at small $\sigma$
are given by
\begin{alignat}{2}
  \label{e.rho(s).asymptotic.alpha<1}
  \hat{\rho}_{\pm}(\sigma)                                & \underset{\sigma \to 0}{\simeq}
  1 - a_{\pm}
  \sigma^{\alpha_{\pm}} + o(\sigma^{\alpha_{\pm}}), \quad & (0 < \alpha_{\pm} < 1)
  \\[0.1cm]
  \label{e.rho(s).asymptotic.alpha>1}
  \hat{\rho}_{\pm}(\sigma)                                & \underset{\sigma \to 0}{\simeq}  1
  - \mu_{\pm} \sigma + a_{\pm} \sigma^{\alpha_{\pm}}  + o(\sigma^{\alpha_{\pm}}),
  \quad                                                   & (1 < \alpha_{\pm} < 2)
\end{alignat}
where $\mu_\pm$ is the mean sojourn time of the state $\pm$.

Now, let us assume that the sojourn-time PDF is given by a power law
distribution. For $0 < \alpha_{\pm} < 1$, the mean sojourn time $\mu_{\pm}$
diverges, and therefore the equilibrium ensemble [Eq.~(\ref{e.eq-ensemble-2})]
does not exist. In contrast, for $1 < \alpha_{\pm} \leq 2$, the equilibrium
ensemble exists and is given by
\begin{equation}
  \label{e.equilibrium.power-law}
  \hat{\rho}^{\pm, \mathrm{eq}} (\sigma)
  =
  \frac {1 - \hat{\rho}^{\pm}(\sigma)}{\mu_{\pm}\sigma}
  \underset{\sigma \to 0}{\simeq} 
  1
  -
  \frac {a_{\pm}}{\mu_{\pm}} \sigma^{\alpha_{\pm}-1},
\end{equation}
where we used Eq.~(\ref{e.rho(s).asymptotic.alpha>1}). The important point is
that the power exponent of $\sigma$ in Eq.~(\ref{e.equilibrium.power-law}) is
smaller by one than that in the sojourn-time PDF
[Eq.~(\ref{e.rho(s).asymptotic.alpha>1})]. This means that the mean initial
sojourn time of the equilibrium ensemble with $1 < \alpha_{\pm} < 2$ diverges.

\section {Simulation setup}\label{sec:app.simulation}

In numerical simulations, we used a simple sojourn-time PDF defined as
\begin{equation}
  \label{e.simulation.rho(tau)}
  \rho_{\pm} (\tau) = \frac {\alpha_{\pm} \tau_0^{\alpha_{\pm}}}{\tau^{1+\alpha_{\pm}}},
  \qquad
  (\tau_0 \leq \tau < \infty)
\end{equation}
where $\tau_0$ is a cutoff time for short trap times; we assume the same cutoff
time for both $\rho_+(\tau)$ and $\rho_-(\tau)$. Then, by comparing
Eq.~(\ref{e.simulation.rho(tau)}) with Eq.~(\ref{e.rho(t)}), it is found that
$a_{\pm}$ in Eq.~(\ref{e.rho(s).power-law(1,2)}) is given by $a_{\pm}=
\tau_0^{\alpha_{\pm}}|\Gamma(1-\alpha_{\pm})|$. For $ \alpha_{\pm} > 1$, the
mean sojourn times $\mu_{\pm}$ are given as
\begin{equation}
  \label{e.app.mu}
  \mu_{\pm} = \frac {\alpha_{\pm} \tau_0}{\alpha_{\pm} - 1}.
\end{equation}


To simulate equilibrium processes, we have to generate initial ensembles which
follow the first sojourn-time PDFs $\rho_{\pm}^{\mathrm{eq}}(t)$
[Eq.~(\ref{e.init-deinsity-equilibrium})]. This can be achieved with a method
presented in Ref.~\cite{miyaguchi13}. First, let $\Theta$ be a random variable
following a PDF
\begin{equation}
  \rho_{\pm}' (\tau)
  =
  \frac {\tau}{\mu_{\pm}} \rho_{\pm} (\tau)
  =
  \frac {(\alpha_{\pm}-1) \tau_0^{\alpha_{\pm}-1}}{\tau^{\alpha_{\pm}}},
  \quad
  (\tau_0 \leq \tau < \infty)
\end{equation}
where Eq.~(\ref{e.app.mu}) is used. Moreover, let $X$ be a random variable,
which follows the uniform PDF on the interval $(0,1)$. Then, a random variable
$X\Theta$ follows the PDF $\rho^{\pm, \mathrm{eq}}(\tau)$ (See
Ref.~\cite{miyaguchi13} for a derivation).


The Langevin equation in Eq.~(\ref{e.ou_proc.w.fluctuating-diffusivity}), or its
discretized form,
\begin{equation}
  d\bm{r}(t)
  =
  - u D(t) \bm{r}(t) dt
  + \sqrt{2 D(t) dt}\, \bm{\xi}(t),
\end{equation}
can be transformed into a dimensionless form with
\begin{equation}
  \tilde{\bm{r}}(t) = \frac {\bm{r}(t)}{\sqrt{D_+ \tau_0}}, \qquad
  \tilde{t} = \frac {t}{\tau_0}, \qquad
  \tilde{u} = u D_+ \tau_0
\end{equation}
The remaining system parameters are $\alpha_{\pm}$, $\tilde{u}$ and the ratio
$D_-/D_+$. In simulations, we set $D_+ = 1.0$ and $\tau_0 = 1$. For the
numerical integration of the Langevin equation, the Euler method is employed
\cite{kloeden11}.


%


\begin{thebibliography}{55}%
\makeatletter
\providecommand \@ifxundefined [1]{%
 \@ifx{#1\undefined}
}%
\providecommand \@ifnum [1]{%
 \ifnum #1\expandafter \@firstoftwo
 \else \expandafter \@secondoftwo
 \fi
}%
\providecommand \@ifx [1]{%
 \ifx #1\expandafter \@firstoftwo
 \else \expandafter \@secondoftwo
 \fi
}%
\providecommand \natexlab [1]{#1}%
%
\providecommand \bibnamefont  [1]{#1}%
\providecommand \bibfnamefont [1]{#1}%
\providecommand \citenamefont [1]{#1}%
\providecommand \href@noop [0]{\@secondoftwo}%
\providecommand \href [0]{\begingroup \@sanitize@url \@href}%
\providecommand \@href[1]{\@@startlink{#1}\@@href}%
\providecommand \@@href[1]{\endgroup#1\@@endlink}%
\providecommand \@sanitize@url [0]{\catcode `\\12\catcode `\$12\catcode
  `\&12\catcode `\#12\catcode `\^12\catcode `\_12\catcode `\%12\relax}%
\providecommand \@@startlink[1]{}%
\providecommand \@@endlink[0]{}%
\providecommand \url  [0]{\begingroup\@sanitize@url \@url }%
\providecommand \@url [1]{\endgroup\@href {#1}{\urlprefix }}%
\providecommand \urlprefix  [0]{URL }%
%
\providecommand \doibase [0]{http://dx.doi.org/}%
\providecommand \selectlanguage [0]{\@gobble}%
\providecommand \bibinfo  [0]{\@secondoftwo}%
\providecommand \bibfield  [0]{\@secondoftwo}%
%
\providecommand \BibitemOpen [0]{}%
%
%
%
\providecommand \BibitemShut  [1]{\csname bibitem#1\endcsname}%
\let\auto@bib@innerbib\@empty
\bibitem [{\citenamefont {Dhont}(1996)}]{dhont96}%
  \BibitemOpen
  \bibfield  {author} {\bibinfo {author} {\bibfnamefont {J.~K.~G.}\
  \bibnamefont {Dhont}},\ }\href@noop {} {\emph {\bibinfo {title} {An
  Introduction to Dynamics of Colloids}}}\ (\bibinfo  {publisher} {Elsevier},\
  \bibinfo {address} {Amsterdam},\ \bibinfo {year} {1996})\BibitemShut
  {NoStop}%
\bibitem [{\citenamefont {Doi}\ and\ \citenamefont {Edwards}(1986)}]{doi86}%
  \BibitemOpen
  \bibfield  {author} {\bibinfo {author} {\bibfnamefont {M.}~\bibnamefont
  {Doi}}\ and\ \bibinfo {author} {\bibfnamefont {S.~F.}\ \bibnamefont
  {Edwards}},\ }\href@noop {} {\emph {\bibinfo {title} {The Theory of Polymer
  Dynamics}}}\ (\bibinfo  {publisher} {Oxford University Press},\ \bibinfo
  {address} {Oxford},\ \bibinfo {year} {1986})\BibitemShut {NoStop}%
\bibitem [{\citenamefont {Bressloff}\ and\ \citenamefont
  {Newby}(2013)}]{bressloff13}%
  \BibitemOpen
  \bibfield  {author} {\bibinfo {author} {\bibfnamefont {P.~C.}\ \bibnamefont
  {Bressloff}}\ and\ \bibinfo {author} {\bibfnamefont {J.~M.}\ \bibnamefont
  {Newby}},\ }\href@noop {} {\bibfield  {journal} {\bibinfo  {journal} {Rev.
  Mod. Phys.}\ }\textbf {\bibinfo {volume} {85}},\ \bibinfo {pages} {135}
  (\bibinfo {year} {2013})}\BibitemShut {NoStop}%
\bibitem [{\citenamefont {Landau}\ and\ \citenamefont
  {Lifshitz}(1987)}]{landau_fluid_mechanics}%
  \BibitemOpen
  \bibfield  {author} {\bibinfo {author} {\bibfnamefont {L.~D.}\ \bibnamefont
  {Landau}}\ and\ \bibinfo {author} {\bibfnamefont {E.}~\bibnamefont
  {Lifshitz}},\ }\href@noop {} {\emph {\bibinfo {title} {Fluid Mechanics}}},\
  \bibinfo {edition} {2nd}\ ed.\ (\bibinfo  {publisher} {Elsevier},\ \bibinfo
  {address} {Oxford},\ \bibinfo {year} {1987})\BibitemShut {NoStop}%
\bibitem [{\citenamefont {Kim}\ and\ \citenamefont {Karrila}(2013)}]{kim13}%
  \BibitemOpen
  \bibfield  {author} {\bibinfo {author} {\bibfnamefont {S.}~\bibnamefont
  {Kim}}\ and\ \bibinfo {author} {\bibfnamefont {S.~J.}\ \bibnamefont
  {Karrila}},\ }\href@noop {} {\emph {\bibinfo {title} {Microhydrodynamics:
  principles and selected applications}}}\ (\bibinfo  {publisher} {Dover},\
  \bibinfo {address} {New York},\ \bibinfo {year} {2013})\BibitemShut {NoStop}%
\bibitem [{\citenamefont {Yamamoto}\ and\ \citenamefont
  {Onuki}(1998{\natexlab{a}})}]{yamamoto98a}%
  \BibitemOpen
  \bibfield  {author} {\bibinfo {author} {\bibfnamefont {R.}~\bibnamefont
  {Yamamoto}}\ and\ \bibinfo {author} {\bibfnamefont {A.}~\bibnamefont
  {Onuki}},\ }\href@noop {} {\bibfield  {journal} {\bibinfo  {journal} {Phys.
  Rev. Lett.}\ }\textbf {\bibinfo {volume} {81}},\ \bibinfo {pages} {4915}
  (\bibinfo {year} {1998}{\natexlab{a}})}\BibitemShut {NoStop}%
\bibitem [{\citenamefont {Yamamoto}\ and\ \citenamefont
  {Onuki}(1998{\natexlab{b}})}]{yamamoto98b}%
  \BibitemOpen
  \bibfield  {author} {\bibinfo {author} {\bibfnamefont {R.}~\bibnamefont
  {Yamamoto}}\ and\ \bibinfo {author} {\bibfnamefont {A.}~\bibnamefont
  {Onuki}},\ }\href@noop {} {\bibfield  {journal} {\bibinfo  {journal} {Phys.
  Rev. E}\ }\textbf {\bibinfo {volume} {58}},\ \bibinfo {pages} {3515}
  (\bibinfo {year} {1998}{\natexlab{b}})}\BibitemShut {NoStop}%
\bibitem [{\citenamefont {Parry}\ \emph {et~al.}(2014)\citenamefont {Parry},
  \citenamefont {Surovtsev}, \citenamefont {Cabeen}, \citenamefont {O'Hern},
  \citenamefont {Dufresne},\ and\ \citenamefont {Jacobs-Wagner}}]{parry14}%
  \BibitemOpen
  \bibfield  {author} {\bibinfo {author} {\bibfnamefont {B.~R.}\ \bibnamefont
  {Parry}}, \bibinfo {author} {\bibfnamefont {I.~V.}\ \bibnamefont
  {Surovtsev}}, \bibinfo {author} {\bibfnamefont {M.~T.}\ \bibnamefont
  {Cabeen}}, \bibinfo {author} {\bibfnamefont {C.~S.}\ \bibnamefont {O'Hern}},
  \bibinfo {author} {\bibfnamefont {E.~R.}\ \bibnamefont {Dufresne}}, \ and\
  \bibinfo {author} {\bibfnamefont {C.}~\bibnamefont {Jacobs-Wagner}},\ }\href
  {\doibase http://dx.doi.org/10.1016/j.cell.2013.11.028} {\bibfield  {journal}
  {\bibinfo  {journal} {Cell}\ }\textbf {\bibinfo {volume} {156}},\ \bibinfo
  {pages} {183 } (\bibinfo {year} {2014})}\BibitemShut {NoStop}%
\bibitem [{\citenamefont {Uneyama}\ \emph {et~al.}(2012)\citenamefont
  {Uneyama}, \citenamefont {Akimoto},\ and\ \citenamefont
  {Miyaguchi}}]{uneyama12}%
  \BibitemOpen
  \bibfield  {author} {\bibinfo {author} {\bibfnamefont {T.}~\bibnamefont
  {Uneyama}}, \bibinfo {author} {\bibfnamefont {T.}~\bibnamefont {Akimoto}}, \
  and\ \bibinfo {author} {\bibfnamefont {T.}~\bibnamefont {Miyaguchi}},\ }\href
  {\doibase 10.1063/1.4752768} {\bibfield  {journal} {\bibinfo  {journal} {J.
  Chem. Phys.}\ }\textbf {\bibinfo {volume} {137}},\ \bibinfo {eid} {114903}
  (\bibinfo {year} {2012})}\BibitemShut {NoStop}%
\bibitem [{\citenamefont {Uneyama}\ \emph {et~al.}(2015)\citenamefont
  {Uneyama}, \citenamefont {Miyaguchi},\ and\ \citenamefont
  {Akimoto}}]{uneyama15}%
  \BibitemOpen
  \bibfield  {author} {\bibinfo {author} {\bibfnamefont {T.}~\bibnamefont
  {Uneyama}}, \bibinfo {author} {\bibfnamefont {T.}~\bibnamefont {Miyaguchi}},
  \ and\ \bibinfo {author} {\bibfnamefont {T.}~\bibnamefont {Akimoto}},\ }\href
  {\doibase 10.1103/PhysRevE.92.032140} {\bibfield  {journal} {\bibinfo
  {journal} {Phys. Rev. E}\ }\textbf {\bibinfo {volume} {92}},\ \bibinfo
  {pages} {032140} (\bibinfo {year} {2015})}\BibitemShut {NoStop}%
\bibitem [{\citenamefont {Miyaguchi}(2017)}]{miyaguchi17}%
  \BibitemOpen
  \bibfield  {author} {\bibinfo {author} {\bibfnamefont {T.}~\bibnamefont
  {Miyaguchi}},\ }\href {\doibase 10.1103/PhysRevE.96.042501} {\bibfield
  {journal} {\bibinfo  {journal} {Phys. Rev. E}\ }\textbf {\bibinfo {volume}
  {96}},\ \bibinfo {pages} {042501} (\bibinfo {year} {2017})}\BibitemShut
  {NoStop}%
\bibitem [{\citenamefont {Chubynsky}\ and\ \citenamefont
  {Slater}(2014)}]{mykyta14}%
  \BibitemOpen
  \bibfield  {author} {\bibinfo {author} {\bibfnamefont {M.~V.}\ \bibnamefont
  {Chubynsky}}\ and\ \bibinfo {author} {\bibfnamefont {G.~W.}\ \bibnamefont
  {Slater}},\ }\href {\doibase 10.1103/PhysRevLett.113.098302} {\bibfield
  {journal} {\bibinfo  {journal} {Phys. Rev. Lett.}\ }\textbf {\bibinfo
  {volume} {113}},\ \bibinfo {pages} {098302} (\bibinfo {year}
  {2014})}\BibitemShut {NoStop}%
\bibitem [{\citenamefont {Manzo}\ \emph {et~al.}(2015)\citenamefont {Manzo},
  \citenamefont {Torreno-Pina}, \citenamefont {Massignan}, \citenamefont
  {Lapeyre}, \citenamefont {Lewenstein},\ and\ \citenamefont
  {Garcia~Parajo}}]{manzo15}%
  \BibitemOpen
  \bibfield  {author} {\bibinfo {author} {\bibfnamefont {C.}~\bibnamefont
  {Manzo}}, \bibinfo {author} {\bibfnamefont {J.~A.}\ \bibnamefont
  {Torreno-Pina}}, \bibinfo {author} {\bibfnamefont {P.}~\bibnamefont
  {Massignan}}, \bibinfo {author} {\bibfnamefont {G.~J.}\ \bibnamefont
  {Lapeyre}}, \bibinfo {author} {\bibfnamefont {M.}~\bibnamefont {Lewenstein}},
  \ and\ \bibinfo {author} {\bibfnamefont {M.~F.}\ \bibnamefont
  {Garcia~Parajo}},\ }\href {\doibase 10.1103/PhysRevX.5.011021} {\bibfield
  {journal} {\bibinfo  {journal} {Phys. Rev. X}\ }\textbf {\bibinfo {volume}
  {5}},\ \bibinfo {pages} {011021} (\bibinfo {year} {2015})}\BibitemShut
  {NoStop}%
\bibitem [{\citenamefont {Chechkin}\ \emph {et~al.}(2017)\citenamefont
  {Chechkin}, \citenamefont {Seno}, \citenamefont {Metzler},\ and\
  \citenamefont {Sokolov}}]{chechkin17}%
  \BibitemOpen
  \bibfield  {author} {\bibinfo {author} {\bibfnamefont {A.~V.}\ \bibnamefont
  {Chechkin}}, \bibinfo {author} {\bibfnamefont {F.}~\bibnamefont {Seno}},
  \bibinfo {author} {\bibfnamefont {R.}~\bibnamefont {Metzler}}, \ and\
  \bibinfo {author} {\bibfnamefont {I.~M.}\ \bibnamefont {Sokolov}},\ }\href
  {\doibase 10.1103/PhysRevX.7.021002} {\bibfield  {journal} {\bibinfo
  {journal} {Phys. Rev. X}\ }\textbf {\bibinfo {volume} {7}},\ \bibinfo {pages}
  {021002} (\bibinfo {year} {2017})}\BibitemShut {NoStop}%
\bibitem [{\citenamefont {Jain}\ and\ \citenamefont
  {Sebastian}(2017)}]{jain17}%
  \BibitemOpen
  \bibfield  {author} {\bibinfo {author} {\bibfnamefont {R.}~\bibnamefont
  {Jain}}\ and\ \bibinfo {author} {\bibfnamefont {K.~L.}\ \bibnamefont
  {Sebastian}},\ }\href@noop {} {\bibfield  {journal} {\bibinfo  {journal} {J.
  Chem. Sci.}\ }\textbf {\bibinfo {volume} {129}},\ \bibinfo {pages} {929}
  (\bibinfo {year} {2017})}\BibitemShut {NoStop}%
\bibitem [{\citenamefont {Golding}\ and\ \citenamefont
  {Cox}(2006)}]{golding06}%
  \BibitemOpen
  \bibfield  {author} {\bibinfo {author} {\bibfnamefont {I.}~\bibnamefont
  {Golding}}\ and\ \bibinfo {author} {\bibfnamefont {E.~C.}\ \bibnamefont
  {Cox}},\ }\href@noop {} {\bibfield  {journal} {\bibinfo  {journal} {Phys.
  Rev. Lett.}\ }\textbf {\bibinfo {volume} {96}},\ \bibinfo {pages} {098102}
  (\bibinfo {year} {2006})}\BibitemShut {NoStop}%
\bibitem [{\citenamefont {Milo}\ and\ \citenamefont {Phillips}(2015)}]{milo15}%
  \BibitemOpen
  \bibfield  {author} {\bibinfo {author} {\bibfnamefont {R.}~\bibnamefont
  {Milo}}\ and\ \bibinfo {author} {\bibfnamefont {R.}~\bibnamefont
  {Phillips}},\ }\href@noop {} {\emph {\bibinfo {title} {Cell biology by the
  numbers}}}\ (\bibinfo  {publisher} {Garland Science},\ \bibinfo {address}
  {New York},\ \bibinfo {year} {2015})\BibitemShut {NoStop}%
\bibitem [{\citenamefont {Uneyama}\ \emph {et~al.}(2019)\citenamefont
  {Uneyama}, \citenamefont {Miyaguchi},\ and\ \citenamefont
  {Akimoto}}]{uneyama19}%
  \BibitemOpen
  \bibfield  {author} {\bibinfo {author} {\bibfnamefont {T.}~\bibnamefont
  {Uneyama}}, \bibinfo {author} {\bibfnamefont {T.}~\bibnamefont {Miyaguchi}},
  \ and\ \bibinfo {author} {\bibfnamefont {T.}~\bibnamefont {Akimoto}},\ }\href
  {\doibase 10.1103/PhysRevE.99.032127} {\bibfield  {journal} {\bibinfo
  {journal} {Phys. Rev. E}\ }\textbf {\bibinfo {volume} {99}},\ \bibinfo
  {pages} {032127} (\bibinfo {year} {2019})}\BibitemShut {NoStop}%
\bibitem [{\citenamefont {Weber}\ \emph {et~al.}(2010)\citenamefont {Weber},
  \citenamefont {Spakowitz},\ and\ \citenamefont {Theriot}}]{weber10}%
  \BibitemOpen
  \bibfield  {author} {\bibinfo {author} {\bibfnamefont {S.~C.}\ \bibnamefont
  {Weber}}, \bibinfo {author} {\bibfnamefont {A.~J.}\ \bibnamefont
  {Spakowitz}}, \ and\ \bibinfo {author} {\bibfnamefont {J.~A.}\ \bibnamefont
  {Theriot}},\ }\href {\doibase 10.1103/PhysRevLett.104.238102} {\bibfield
  {journal} {\bibinfo  {journal} {Phys. Rev. Lett.}\ }\textbf {\bibinfo
  {volume} {104}},\ \bibinfo {pages} {238102} (\bibinfo {year}
  {2010})}\BibitemShut {NoStop}%
\bibitem [{\citenamefont {Weigel}\ \emph {et~al.}(2011)\citenamefont {Weigel},
  \citenamefont {Simon}, \citenamefont {Tamkun},\ and\ \citenamefont
  {Krapf}}]{weigel11}%
  \BibitemOpen
  \bibfield  {author} {\bibinfo {author} {\bibfnamefont {A.~V.}\ \bibnamefont
  {Weigel}}, \bibinfo {author} {\bibfnamefont {B.}~\bibnamefont {Simon}},
  \bibinfo {author} {\bibfnamefont {M.~M.}\ \bibnamefont {Tamkun}}, \ and\
  \bibinfo {author} {\bibfnamefont {D.}~\bibnamefont {Krapf}},\ }\href@noop {}
  {\bibfield  {journal} {\bibinfo  {journal} {Proc. Natl. Acad. Sci. U.S.A}\
  }\textbf {\bibinfo {volume} {108}},\ \bibinfo {pages} {6438} (\bibinfo {year}
  {2011})}\BibitemShut {NoStop}%
\bibitem [{\citenamefont {Jeon}\ \emph {et~al.}(2011)\citenamefont {Jeon},
  \citenamefont {Tejedor}, \citenamefont {Burov}, \citenamefont {Barkai},
  \citenamefont {Selhuber-Unkel}, \citenamefont {Berg-S\o{}rensen},
  \citenamefont {Oddershede},\ and\ \citenamefont {Metzler}}]{jeon11}%
  \BibitemOpen
  \bibfield  {author} {\bibinfo {author} {\bibfnamefont {J.-H.}\ \bibnamefont
  {Jeon}}, \bibinfo {author} {\bibfnamefont {V.}~\bibnamefont {Tejedor}},
  \bibinfo {author} {\bibfnamefont {S.}~\bibnamefont {Burov}}, \bibinfo
  {author} {\bibfnamefont {E.}~\bibnamefont {Barkai}}, \bibinfo {author}
  {\bibfnamefont {C.}~\bibnamefont {Selhuber-Unkel}}, \bibinfo {author}
  {\bibfnamefont {K.}~\bibnamefont {Berg-S\o{}rensen}}, \bibinfo {author}
  {\bibfnamefont {L.}~\bibnamefont {Oddershede}}, \ and\ \bibinfo {author}
  {\bibfnamefont {R.}~\bibnamefont {Metzler}},\ }\href {\doibase
  10.1103/PhysRevLett.106.048103} {\bibfield  {journal} {\bibinfo  {journal}
  {Phys. Rev. Lett.}\ }\textbf {\bibinfo {volume} {106}},\ \bibinfo {pages}
  {048103} (\bibinfo {year} {2011})}\BibitemShut {NoStop}%
\bibitem [{\citenamefont {Burov}\ \emph {et~al.}(2011)\citenamefont {Burov},
  \citenamefont {Jeon}, \citenamefont {Metzler},\ and\ \citenamefont
  {Barkai}}]{burov11}%
  \BibitemOpen
  \bibfield  {author} {\bibinfo {author} {\bibfnamefont {S.}~\bibnamefont
  {Burov}}, \bibinfo {author} {\bibfnamefont {J.}~\bibnamefont {Jeon}},
  \bibinfo {author} {\bibfnamefont {R.}~\bibnamefont {Metzler}}, \ and\
  \bibinfo {author} {\bibfnamefont {E.}~\bibnamefont {Barkai}},\ }\href
  {\doibase 10.1039/c0cp01879a} {\bibfield  {journal} {\bibinfo  {journal}
  {Phys. Chem. Chem. Phys.}\ }\textbf {\bibinfo {volume} {13}},\ \bibinfo
  {pages} {1800} (\bibinfo {year} {2011})}\BibitemShut {NoStop}%
\bibitem [{\citenamefont {Tabei}\ \emph {et~al.}(2013)\citenamefont {Tabei},
  \citenamefont {Burov}, \citenamefont {Kim}, \citenamefont {Kuznetsov},
  \citenamefont {Huynh}, \citenamefont {Jureller}, \citenamefont
  {H.~Philipson}, \citenamefont {Dinner},\ and\ \citenamefont
  {Scherer}}]{tabei13}%
  \BibitemOpen
  \bibfield  {author} {\bibinfo {author} {\bibfnamefont {S.~M.~A.}\
  \bibnamefont {Tabei}}, \bibinfo {author} {\bibfnamefont {S.}~\bibnamefont
  {Burov}}, \bibinfo {author} {\bibfnamefont {H.~Y.}\ \bibnamefont {Kim}},
  \bibinfo {author} {\bibfnamefont {A.}~\bibnamefont {Kuznetsov}}, \bibinfo
  {author} {\bibfnamefont {T.}~\bibnamefont {Huynh}}, \bibinfo {author}
  {\bibfnamefont {J.}~\bibnamefont {Jureller}}, \bibinfo {author}
  {\bibfnamefont {L.}~\bibnamefont {H.~Philipson}}, \bibinfo {author}
  {\bibfnamefont {A.~R.}\ \bibnamefont {Dinner}}, \ and\ \bibinfo {author}
  {\bibfnamefont {N.~F.}\ \bibnamefont {Scherer}},\ }\href {\doibase
  10.1073/pnas.0508366103} {\bibfield  {journal} {\bibinfo  {journal} {Proc.
  Natl. Acad. Sci. U.S.A}\ }\textbf {\bibinfo {volume} {110}},\ \bibinfo
  {pages} {4911} (\bibinfo {year} {2013})}\BibitemShut {NoStop}%
\bibitem [{\citenamefont {Yamamoto}\ \emph {et~al.}(2014)\citenamefont
  {Yamamoto}, \citenamefont {Akimoto}, \citenamefont {Yasui},\ and\
  \citenamefont {Yasuoka}}]{yamamoto14}%
  \BibitemOpen
  \bibfield  {author} {\bibinfo {author} {\bibfnamefont {E.}~\bibnamefont
  {Yamamoto}}, \bibinfo {author} {\bibfnamefont {T.}~\bibnamefont {Akimoto}},
  \bibinfo {author} {\bibfnamefont {M.}~\bibnamefont {Yasui}}, \ and\ \bibinfo
  {author} {\bibfnamefont {K.}~\bibnamefont {Yasuoka}},\ }\href@noop {}
  {\bibfield  {journal} {\bibinfo  {journal} {Sci. Rep.}\ }\textbf {\bibinfo
  {volume} {4}},\ \bibinfo {pages} {4720} (\bibinfo {year} {2014})}\BibitemShut
  {NoStop}%
\bibitem [{\citenamefont {He}\ \emph {et~al.}(2008)\citenamefont {He},
  \citenamefont {Burov}, \citenamefont {Metzler},\ and\ \citenamefont
  {Barkai}}]{he08}%
  \BibitemOpen
  \bibfield  {author} {\bibinfo {author} {\bibfnamefont {Y.}~\bibnamefont
  {He}}, \bibinfo {author} {\bibfnamefont {S.}~\bibnamefont {Burov}}, \bibinfo
  {author} {\bibfnamefont {R.}~\bibnamefont {Metzler}}, \ and\ \bibinfo
  {author} {\bibfnamefont {E.}~\bibnamefont {Barkai}},\ }\href@noop {}
  {\bibfield  {journal} {\bibinfo  {journal} {Phys. Rev. Lett.}\ }\textbf
  {\bibinfo {volume} {101}},\ \bibinfo {pages} {058101} (\bibinfo {year}
  {2008})}\BibitemShut {NoStop}%
\bibitem [{\citenamefont {Meroz}\ \emph {et~al.}(2010)\citenamefont {Meroz},
  \citenamefont {Sokolov},\ and\ \citenamefont {Klafter}}]{meroz10}%
  \BibitemOpen
  \bibfield  {author} {\bibinfo {author} {\bibfnamefont {Y.}~\bibnamefont
  {Meroz}}, \bibinfo {author} {\bibfnamefont {I.~M.}\ \bibnamefont {Sokolov}},
  \ and\ \bibinfo {author} {\bibfnamefont {J.}~\bibnamefont {Klafter}},\
  }\href@noop {} {\bibfield  {journal} {\bibinfo  {journal} {Phys. Rev. E}\
  }\textbf {\bibinfo {volume} {81}},\ \bibinfo {pages} {010101(R)} (\bibinfo
  {year} {2010})}\BibitemShut {NoStop}%
\bibitem [{\citenamefont {Miyaguchi}\ and\ \citenamefont
  {Akimoto}(2011)}]{miyaguchi11b}%
  \BibitemOpen
  \bibfield  {author} {\bibinfo {author} {\bibfnamefont {T.}~\bibnamefont
  {Miyaguchi}}\ and\ \bibinfo {author} {\bibfnamefont {T.}~\bibnamefont
  {Akimoto}},\ }\href {\doibase 10.1103/PhysRevE.83.031926} {\bibfield
  {journal} {\bibinfo  {journal} {Phys. Rev. E}\ }\textbf {\bibinfo {volume}
  {83}},\ \bibinfo {pages} {031926} (\bibinfo {year} {2011})}\BibitemShut
  {NoStop}%
\bibitem [{\citenamefont {Leith}\ \emph {et~al.}(2012)\citenamefont {Leith},
  \citenamefont {Tafvizi}, \citenamefont {Huang}, \citenamefont {Uspal},
  \citenamefont {Doyle}, \citenamefont {Fersht}, \citenamefont {Mirny},\ and\
  \citenamefont {van Oijen}}]{leith12}%
  \BibitemOpen
  \bibfield  {author} {\bibinfo {author} {\bibfnamefont {J.~S.}\ \bibnamefont
  {Leith}}, \bibinfo {author} {\bibfnamefont {A.}~\bibnamefont {Tafvizi}},
  \bibinfo {author} {\bibfnamefont {F.}~\bibnamefont {Huang}}, \bibinfo
  {author} {\bibfnamefont {W.~E.}\ \bibnamefont {Uspal}}, \bibinfo {author}
  {\bibfnamefont {P.~S.}\ \bibnamefont {Doyle}}, \bibinfo {author}
  {\bibfnamefont {A.~R.}\ \bibnamefont {Fersht}}, \bibinfo {author}
  {\bibfnamefont {L.~A.}\ \bibnamefont {Mirny}}, \ and\ \bibinfo {author}
  {\bibfnamefont {A.~M.}\ \bibnamefont {van Oijen}},\ }\href {\doibase
  10.1073/pnas.1120452109} {\bibfield  {journal} {\bibinfo  {journal} {Proc.
  Natl. Acad. Sci. U.S.A}\ }\textbf {\bibinfo {volume} {109}},\ \bibinfo
  {pages} {16552} (\bibinfo {year} {2012})}\BibitemShut {NoStop}%
\bibitem [{\citenamefont {Detcheverry}(2017)}]{detcheverry17}%
  \BibitemOpen
  \bibfield  {author} {\bibinfo {author} {\bibfnamefont {F.}~\bibnamefont
  {Detcheverry}},\ }\href@noop {} {\bibfield  {journal} {\bibinfo  {journal}
  {Phys. Rev. E}\ }\textbf {\bibinfo {volume} {96}},\ \bibinfo {pages} {012415}
  (\bibinfo {year} {2017})}\BibitemShut {NoStop}%
\bibitem [{\citenamefont {Cox}(1962)}]{cox62}%
  \BibitemOpen
  \bibfield  {author} {\bibinfo {author} {\bibfnamefont {D.~R.}\ \bibnamefont
  {Cox}},\ }\href@noop {} {\emph {\bibinfo {title} {Renewal Theory}}}\
  (\bibinfo  {publisher} {Methuen},\ \bibinfo {address} {London},\ \bibinfo
  {year} {1962})\BibitemShut {NoStop}%
\bibitem [{\citenamefont {Goychuk}\ and\ \citenamefont
  {H\"anggi}(2003)}]{goychuk03}%
  \BibitemOpen
  \bibfield  {author} {\bibinfo {author} {\bibfnamefont {I.}~\bibnamefont
  {Goychuk}}\ and\ \bibinfo {author} {\bibfnamefont {P.}~\bibnamefont
  {H\"anggi}},\ }\href {\doibase 10.1103/PhysRevLett.91.070601} {\bibfield
  {journal} {\bibinfo  {journal} {Phys. Rev. Lett.}\ }\textbf {\bibinfo
  {volume} {91}},\ \bibinfo {pages} {070601} (\bibinfo {year}
  {2003})}\BibitemShut {NoStop}%
\bibitem [{\citenamefont {Akimoto}\ and\ \citenamefont
  {Seki}(2015)}]{akimoto15}%
  \BibitemOpen
  \bibfield  {author} {\bibinfo {author} {\bibfnamefont {T.}~\bibnamefont
  {Akimoto}}\ and\ \bibinfo {author} {\bibfnamefont {K.}~\bibnamefont {Seki}},\
  }\href {\doibase 10.1103/PhysRevE.92.022114} {\bibfield  {journal} {\bibinfo
  {journal} {Phys. Rev. E}\ }\textbf {\bibinfo {volume} {92}},\ \bibinfo
  {pages} {022114} (\bibinfo {year} {2015})}\BibitemShut {NoStop}%
\bibitem [{\citenamefont {Miyaguchi}\ \emph {et~al.}(2016)\citenamefont
  {Miyaguchi}, \citenamefont {Akimoto},\ and\ \citenamefont
  {Yamamoto}}]{miyaguchi16}%
  \BibitemOpen
  \bibfield  {author} {\bibinfo {author} {\bibfnamefont {T.}~\bibnamefont
  {Miyaguchi}}, \bibinfo {author} {\bibfnamefont {T.}~\bibnamefont {Akimoto}},
  \ and\ \bibinfo {author} {\bibfnamefont {E.}~\bibnamefont {Yamamoto}},\
  }\href {\doibase 10.1103/PhysRevE.94.012109} {\bibfield  {journal} {\bibinfo
  {journal} {Phys. Rev. E}\ }\textbf {\bibinfo {volume} {94}},\ \bibinfo
  {pages} {012109} (\bibinfo {year} {2016})}\BibitemShut {NoStop}%
\bibitem [{\citenamefont {Odagaki}\ and\ \citenamefont
  {Hiwatari}(1990)}]{odagaki90}%
  \BibitemOpen
  \bibfield  {author} {\bibinfo {author} {\bibfnamefont {T.}~\bibnamefont
  {Odagaki}}\ and\ \bibinfo {author} {\bibfnamefont {Y.}~\bibnamefont
  {Hiwatari}},\ }\href {\doibase 10.1103/PhysRevA.41.929} {\bibfield  {journal}
  {\bibinfo  {journal} {Phys. Rev. A}\ }\textbf {\bibinfo {volume} {41}},\
  \bibinfo {pages} {929} (\bibinfo {year} {1990})}\BibitemShut {NoStop}%
\bibitem [{\citenamefont {Doliwa}\ and\ \citenamefont
  {Heuer}(2003)}]{doliwa03}%
  \BibitemOpen
  \bibfield  {author} {\bibinfo {author} {\bibfnamefont {B.}~\bibnamefont
  {Doliwa}}\ and\ \bibinfo {author} {\bibfnamefont {A.}~\bibnamefont {Heuer}},\
  }\href {\doibase 10.1103/PhysRevE.67.030501} {\bibfield  {journal} {\bibinfo
  {journal} {Phys. Rev. E}\ }\textbf {\bibinfo {volume} {67}},\ \bibinfo
  {pages} {030501(R)} (\bibinfo {year} {2003})}\BibitemShut {NoStop}%
\bibitem [{\citenamefont {Helfferich}\ \emph {et~al.}(2014)\citenamefont
  {Helfferich}, \citenamefont {Ziebert}, \citenamefont {Frey}, \citenamefont
  {Meyer}, \citenamefont {Farago}, \citenamefont {Blumen},\ and\ \citenamefont
  {Baschnagel}}]{helfferich14a}%
  \BibitemOpen
  \bibfield  {author} {\bibinfo {author} {\bibfnamefont {J.}~\bibnamefont
  {Helfferich}}, \bibinfo {author} {\bibfnamefont {F.}~\bibnamefont {Ziebert}},
  \bibinfo {author} {\bibfnamefont {S.}~\bibnamefont {Frey}}, \bibinfo {author}
  {\bibfnamefont {H.}~\bibnamefont {Meyer}}, \bibinfo {author} {\bibfnamefont
  {J.}~\bibnamefont {Farago}}, \bibinfo {author} {\bibfnamefont
  {A.}~\bibnamefont {Blumen}}, \ and\ \bibinfo {author} {\bibfnamefont
  {J.}~\bibnamefont {Baschnagel}},\ }\href {\doibase
  10.1103/PhysRevE.89.042603} {\bibfield  {journal} {\bibinfo  {journal} {Phys.
  Rev. E}\ }\textbf {\bibinfo {volume} {89}},\ \bibinfo {pages} {042603}
  (\bibinfo {year} {2014})}\BibitemShut {NoStop}%
\bibitem [{\citenamefont {Sekimoto}(2010)}]{sekimoto10}%
  \BibitemOpen
  \bibfield  {author} {\bibinfo {author} {\bibfnamefont {K.}~\bibnamefont
  {Sekimoto}},\ }\href@noop {} {\emph {\bibinfo {title} {Stochastic
  Energetics}}}\ (\bibinfo  {publisher} {Springer},\ \bibinfo {address}
  {Berlin},\ \bibinfo {year} {2010})\BibitemShut {NoStop}%
\bibitem [{\citenamefont {Hansen}\ and\ \citenamefont
  {McDonald}(1990)}]{hansen90}%
  \BibitemOpen
  \bibfield  {author} {\bibinfo {author} {\bibfnamefont {J.-P.}\ \bibnamefont
  {Hansen}}\ and\ \bibinfo {author} {\bibfnamefont {I.~R.}\ \bibnamefont
  {McDonald}},\ }\href@noop {} {\emph {\bibinfo {title} {Theory of Simple
  Liquids}}}\ (\bibinfo  {publisher} {Elsevier},\ \bibinfo {address} {New
  York},\ \bibinfo {year} {1990})\BibitemShut {NoStop}%
\bibitem [{\citenamefont {Kob}\ and\ \citenamefont {Andersen}(1995)}]{kob95}%
  \BibitemOpen
  \bibfield  {author} {\bibinfo {author} {\bibfnamefont {W.}~\bibnamefont
  {Kob}}\ and\ \bibinfo {author} {\bibfnamefont {H.~C.}\ \bibnamefont
  {Andersen}},\ }\href {\doibase 10.1103/PhysRevE.51.4626} {\bibfield
  {journal} {\bibinfo  {journal} {Phys. Rev. E}\ }\textbf {\bibinfo {volume}
  {51}},\ \bibinfo {pages} {4626} (\bibinfo {year} {1995})}\BibitemShut
  {NoStop}%
\bibitem [{\citenamefont {Berne}\ and\ \citenamefont {Pecora}(2000)}]{berne00}%
  \BibitemOpen
  \bibfield  {author} {\bibinfo {author} {\bibfnamefont {B.~J.}\ \bibnamefont
  {Berne}}\ and\ \bibinfo {author} {\bibfnamefont {R.}~\bibnamefont {Pecora}},\
  }\href@noop {} {\emph {\bibinfo {title} {Dynamic Light Scattering}}}\
  (\bibinfo  {publisher} {Dover},\ \bibinfo {address} {New York},\ \bibinfo
  {year} {2000})\BibitemShut {NoStop}%
\bibitem [{\citenamefont {Godr{\`e}che}\ and\ \citenamefont
  {Luck}(2001)}]{godrche01}%
  \BibitemOpen
  \bibfield  {author} {\bibinfo {author} {\bibfnamefont {C.}~\bibnamefont
  {Godr{\`e}che}}\ and\ \bibinfo {author} {\bibfnamefont {J.~M.}\ \bibnamefont
  {Luck}},\ }\href@noop {} {\bibfield  {journal} {\bibinfo  {journal} {J. Stat.
  Phys.}\ }\textbf {\bibinfo {volume} {104}},\ \bibinfo {pages} {489} (\bibinfo
  {year} {2001})}\BibitemShut {NoStop}%
\bibitem [{\citenamefont {Miyaguchi}\ and\ \citenamefont
  {Akimoto}(2013)}]{miyaguchi13}%
  \BibitemOpen
  \bibfield  {author} {\bibinfo {author} {\bibfnamefont {T.}~\bibnamefont
  {Miyaguchi}}\ and\ \bibinfo {author} {\bibfnamefont {T.}~\bibnamefont
  {Akimoto}},\ }\href@noop {} {\bibfield  {journal} {\bibinfo  {journal} {Phys.
  Rev. E}\ }\textbf {\bibinfo {volume} {87}},\ \bibinfo {pages} {032130}
  (\bibinfo {year} {2013})}\BibitemShut {NoStop}%
\bibitem [{\citenamefont {Akimoto}\ and\ \citenamefont
  {Aizawa}(2007)}]{akimoto07}%
  \BibitemOpen
  \bibfield  {author} {\bibinfo {author} {\bibfnamefont {T.}~\bibnamefont
  {Akimoto}}\ and\ \bibinfo {author} {\bibfnamefont {Y.}~\bibnamefont
  {Aizawa}},\ }\href@noop {} {\bibfield  {journal} {\bibinfo  {journal} {J
  Korean Phys. Soc.}\ }\textbf {\bibinfo {volume} {50}},\ \bibinfo {pages}
  {254} (\bibinfo {year} {2007})}\BibitemShut {NoStop}%
\bibitem [{\citenamefont {Akimoto}\ \emph {et~al.}(2018)\citenamefont
  {Akimoto}, \citenamefont {Cherstvy},\ and\ \citenamefont
  {Metzler}}]{akimoto18}%
  \BibitemOpen
  \bibfield  {author} {\bibinfo {author} {\bibfnamefont {T.}~\bibnamefont
  {Akimoto}}, \bibinfo {author} {\bibfnamefont {A.~G.}\ \bibnamefont
  {Cherstvy}}, \ and\ \bibinfo {author} {\bibfnamefont {R.}~\bibnamefont
  {Metzler}},\ }\href {\doibase 10.1103/PhysRevE.98.022105} {\bibfield
  {journal} {\bibinfo  {journal} {Phys. Rev. E}\ }\textbf {\bibinfo {volume}
  {98}},\ \bibinfo {pages} {022105} (\bibinfo {year} {2018})}\BibitemShut
  {NoStop}%
\bibitem [{\citenamefont {Cherstvy}\ \emph {et~al.}(2013)\citenamefont
  {Cherstvy}, \citenamefont {Chechkin},\ and\ \citenamefont
  {Metzler}}]{cherstvy13}%
  \BibitemOpen
  \bibfield  {author} {\bibinfo {author} {\bibfnamefont {A.~G.}\ \bibnamefont
  {Cherstvy}}, \bibinfo {author} {\bibfnamefont {A.~V.}\ \bibnamefont
  {Chechkin}}, \ and\ \bibinfo {author} {\bibfnamefont {R.}~\bibnamefont
  {Metzler}},\ }\href {http://stacks.iop.org/1367-2630/15/i=8/a=083039}
  {\bibfield  {journal} {\bibinfo  {journal} {New J. Phys.}\ }\textbf {\bibinfo
  {volume} {15}},\ \bibinfo {pages} {083039} (\bibinfo {year}
  {2013})}\BibitemShut {NoStop}%
\bibitem [{\citenamefont {Cherstvy}\ and\ \citenamefont
  {Metzler}(2016)}]{cherstvy16}%
  \BibitemOpen
  \bibfield  {author} {\bibinfo {author} {\bibfnamefont {A.~G.}\ \bibnamefont
  {Cherstvy}}\ and\ \bibinfo {author} {\bibfnamefont {R.}~\bibnamefont
  {Metzler}},\ }\href@noop {} {\bibfield  {journal} {\bibinfo  {journal} {Phys.
  Chem. Chem. Phys.}\ }\textbf {\bibinfo {volume} {18}},\ \bibinfo {pages}
  {23840} (\bibinfo {year} {2016})}\BibitemShut {NoStop}%
\bibitem [{\citenamefont {Metzler}\ \emph {et~al.}(2014)\citenamefont
  {Metzler}, \citenamefont {Jeon}, \citenamefont {Cherstvy},\ and\
  \citenamefont {Barkai}}]{metzler14}%
  \BibitemOpen
  \bibfield  {author} {\bibinfo {author} {\bibfnamefont {R.}~\bibnamefont
  {Metzler}}, \bibinfo {author} {\bibfnamefont {J.-H.}\ \bibnamefont {Jeon}},
  \bibinfo {author} {\bibfnamefont {A.~G.}\ \bibnamefont {Cherstvy}}, \ and\
  \bibinfo {author} {\bibfnamefont {E.}~\bibnamefont {Barkai}},\ }\href
  {\doibase 10.1039/C4CP03465A} {\bibfield  {journal} {\bibinfo  {journal}
  {Phys. Chem. Chem. Phys.}\ }\textbf {\bibinfo {volume} {16}},\ \bibinfo
  {pages} {24128} (\bibinfo {year} {2014})}\BibitemShut {NoStop}%
\bibitem [{\citenamefont {Palmer}\ \emph {et~al.}(1984)\citenamefont {Palmer},
  \citenamefont {Stein}, \citenamefont {Abrahams},\ and\ \citenamefont
  {Anderson}}]{palmer84}%
  \BibitemOpen
  \bibfield  {author} {\bibinfo {author} {\bibfnamefont {R.~G.}\ \bibnamefont
  {Palmer}}, \bibinfo {author} {\bibfnamefont {D.~L.}\ \bibnamefont {Stein}},
  \bibinfo {author} {\bibfnamefont {E.}~\bibnamefont {Abrahams}}, \ and\
  \bibinfo {author} {\bibfnamefont {P.~W.}\ \bibnamefont {Anderson}},\ }\href
  {\doibase 10.1103/PhysRevLett.53.958} {\bibfield  {journal} {\bibinfo
  {journal} {Phys. Rev. Lett.}\ }\textbf {\bibinfo {volume} {53}},\ \bibinfo
  {pages} {958} (\bibinfo {year} {1984})}\BibitemShut {NoStop}%
\bibitem [{\citenamefont {Ngai}\ and\ \citenamefont {Rendell}(1993)}]{ngai93}%
  \BibitemOpen
  \bibfield  {author} {\bibinfo {author} {\bibfnamefont {K.}~\bibnamefont
  {Ngai}}\ and\ \bibinfo {author} {\bibfnamefont {R.}~\bibnamefont {Rendell}},\
  }\href {\doibase https://doi.org/10.1016/0167-7322(93)80027-S} {\bibfield
  {journal} {\bibinfo  {journal} {J. Mol. Liquid}\ }\textbf {\bibinfo {volume}
  {56}},\ \bibinfo {pages} {199 } (\bibinfo {year} {1993})}\BibitemShut
  {NoStop}%
\bibitem [{\citenamefont {Phillips}(1996)}]{phillips96}%
  \BibitemOpen
  \bibfield  {author} {\bibinfo {author} {\bibfnamefont {J.}~\bibnamefont
  {Phillips}},\ }\href@noop {} {\bibfield  {journal} {\bibinfo  {journal} {Rep.
  Prog. Phys.}\ }\textbf {\bibinfo {volume} {59}},\ \bibinfo {pages} {1133}
  (\bibinfo {year} {1996})}\BibitemShut {NoStop}%
\bibitem [{\citenamefont {Metzler}\ and\ \citenamefont
  {Klafter}(2000)}]{metzler00}%
  \BibitemOpen
  \bibfield  {author} {\bibinfo {author} {\bibfnamefont {R.}~\bibnamefont
  {Metzler}}\ and\ \bibinfo {author} {\bibfnamefont {J.}~\bibnamefont
  {Klafter}},\ }\href {\doibase 10.1016/S0370-1573(00)00070-3} {\bibfield
  {journal} {\bibinfo  {journal} {Phys. Rep.}\ }\textbf {\bibinfo {volume}
  {339}},\ \bibinfo {pages} {1} (\bibinfo {year} {2000})}\BibitemShut {NoStop}%
\bibitem [{\citenamefont {Machta}(1985)}]{machta85}%
  \BibitemOpen
  \bibfield  {author} {\bibinfo {author} {\bibfnamefont {J.}~\bibnamefont
  {Machta}},\ }\href@noop {} {\bibfield  {journal} {\bibinfo  {journal}
  {J.~Phys. A}\ }\textbf {\bibinfo {volume} {18}},\ \bibinfo {pages} {L531}
  (\bibinfo {year} {1985})}\BibitemShut {NoStop}%
\bibitem [{\citenamefont {Chaudhuri}\ \emph {et~al.}(2007)\citenamefont
  {Chaudhuri}, \citenamefont {Berthier},\ and\ \citenamefont
  {Kob}}]{chaudhuri07}%
  \BibitemOpen
  \bibfield  {author} {\bibinfo {author} {\bibfnamefont {P.}~\bibnamefont
  {Chaudhuri}}, \bibinfo {author} {\bibfnamefont {L.}~\bibnamefont {Berthier}},
  \ and\ \bibinfo {author} {\bibfnamefont {W.}~\bibnamefont {Kob}},\
  }\href@noop {} {\bibfield  {journal} {\bibinfo  {journal} {Phys. Rev. Lett.}\
  }\textbf {\bibinfo {volume} {99}},\ \bibinfo {pages} {060604} (\bibinfo
  {year} {2007})}\BibitemShut {NoStop}%
\bibitem [{\citenamefont {Helfferich}\ \emph {et~al.}(2018)\citenamefont
  {Helfferich}, \citenamefont {Brisch}, \citenamefont {Meyer}, \citenamefont
  {Benzerara}, \citenamefont {Ziebert}, \citenamefont {Farago},\ and\
  \citenamefont {Baschnagel}}]{helfferich18}%
  \BibitemOpen
  \bibfield  {author} {\bibinfo {author} {\bibfnamefont {J.}~\bibnamefont
  {Helfferich}}, \bibinfo {author} {\bibfnamefont {J.}~\bibnamefont {Brisch}},
  \bibinfo {author} {\bibfnamefont {H.}~\bibnamefont {Meyer}}, \bibinfo
  {author} {\bibfnamefont {O.}~\bibnamefont {Benzerara}}, \bibinfo {author}
  {\bibfnamefont {F.}~\bibnamefont {Ziebert}}, \bibinfo {author} {\bibfnamefont
  {J.}~\bibnamefont {Farago}}, \ and\ \bibinfo {author} {\bibfnamefont
  {J.}~\bibnamefont {Baschnagel}},\ }\href@noop {} {\bibfield  {journal}
  {\bibinfo  {journal} {Eur. Phys. J. E}\ }\textbf {\bibinfo {volume} {41}},\
  \bibinfo {pages} {71} (\bibinfo {year} {2018})}\BibitemShut {NoStop}%
\bibitem [{\citenamefont {Kloeden}\ and\ \citenamefont
  {Platen}(2011)}]{kloeden11}%
  \BibitemOpen
  \bibfield  {author} {\bibinfo {author} {\bibfnamefont {P.~E.}\ \bibnamefont
  {Kloeden}}\ and\ \bibinfo {author} {\bibfnamefont {E.}~\bibnamefont
  {Platen}},\ }\href@noop {} {\emph {\bibinfo {title} {Numerical Solution of
  Stochastic Differential Equations}}}\ (\bibinfo  {publisher} {Springer},\
  \bibinfo {address} {Berlin},\ \bibinfo {year} {2011})\BibitemShut {NoStop}%
\end{thebibliography}
\end {document}